\documentclass[fleqn,usenatbib]{mnras}

\usepackage{newtxtext,newtxmath}
\usepackage[T1]{fontenc}

\DeclareRobustCommand{\VAN}[3]{#2}
\let\VANthebibliography\thebibliography
\def\thebibliography{\DeclareRobustCommand{\VAN}[3]{##3}\VANthebibliography}

\usepackage{graphicx}
\usepackage{array}
\usepackage{amsmath}
\usepackage{mathtools}
\usepackage{gensymb}
\usepackage{hhline}
\usepackage{float}
\usepackage{longtable,booktabs,etoolbox,multirow,tabularx}
\usepackage{CJK}
\usepackage[whole]{bxcjkjatype} 
\usepackage{CJKutf8}
\title[Geometry of Type Ia SN\,2019np]{The Core Normal Type Ia Supernova 2019np -- An Overall Spherical Explosion with an Aspherical Surface Layer and an Aspherical $^{56}$Ni Core
\thanks{Based on observations collected at the European Southern Observatory under ESO program 0102.D-0528.
}
}
\author[P.\ Hoeflich et al.]
{Peter Hoeflich$^{1,\thanks{E-mail: phoeflich@fsu.edu}}$,
Yi Yang\begin{CJK*}{UTF8}{gbsn}
(杨轶)
\end{CJK*}$^{2,3,\mathsection,\thanks{E-mail: yiyangtamu@gmail.com}}$, 
Dietrich Baade$^{4,\thanks{E-mail: dbaade@eso.org}}$,
Aleksandar Cikota$^{5,6}$,
Justyn R. Maund$^{7}$,\
\newauthor
Divya Mishra$^{8}$,
Ferdinando Patat$^{4}$,
Kishore C. Patra$^{2,+}$, 
Lifan Wang$^{8}$, 
J. Craig Wheeler$^{9}$,\
\newauthor
Alexei V. Filippenko$^{2}$,
Avishay Gal-Yam$^{10}$,
Steve Schulze$^{11}$
\\
\\
$^{1}$Department of Physics, Florida State University, Tallahassee, Florida 32306-4350, USA \\
$^{2}$Department of Astronomy, University of California, Berkeley, CA 94720-3411, USA \\
$^{3}$Department of Particle Physics and Astrophysics, Weizmann Institute of Science, Rehovot 76100, Israel \\
$^{4}$European Organisation for Astronomical Research in the Southern Hemisphere (ESO), Karl-Schwarzschild-Str.\ 2, 85748 Garching b.\ M{\"u}nchen, Germany \\
$^{5}$European Organisation for Astronomical Research in the Southern Hemisphere (ESO), Alonso de Cordova 3107, Vitacura, Casilla 19001, Santiago de Chile, Chile \\
$^{6}$ Gemini Observatory/NSF's NOIRLab, Casilla 603, La Serena, Chile\\
$^{7}$Department of Physics and Astronomy, University of Sheffield, Hicks Building, Hounsfield Road, Sheffield S3 7RH, UK\\
$^{8}$George P.\ and Cynthia Woods Mitchell Institute for Fundamental Physics $\&$ Astronomy, Texas A.$\&$M. University, 4242 TAMU, College Station, TX 77843, USA \\
$^{9}$Department of Astronomy, University of Texas, Austin, TX 78712, USA \\
$^{10}$Department of Particle Physics and Astrophysics, Weizmann Institute of Science, 76100 Rehovot, Israel
\\
$^{11}$The Oskar Klein Centre, Department of Astronomy and Department of Physics, Stockholm 
University, AlbaNova, SE-106 91 Stockholm, Sweden
\\
$^{\mathsection}$Bengier-Winslow-Robertson Fellow \\
$^{+}$Nagaraj-Noll-Otellini Graduate Fellow \\
}

\date{Accepted 1/12/23; Received 12/9/22; in original from 11/09/22}

\pubyear{2022}

\begin{document}
\label{firstpage}
\pagerange{\pageref{firstpage}--\pageref{lastpage}}
\maketitle

% Abstract of the paper
\begin{abstract}
Optical spectropolarimetry of the normal thermonuclear supernova(SN) 2019np from $-$14.5 to $+$14.5 days relative to $B$-band maximum detected an intrinsic continuum polarization ($p^{\rm cont}$) of 0.21\%$\pm$0.09\% at the first epoch.   Between days $-$11.5 and +0.5, $p^{\rm cont}$ remained $\sim$0 and by day $+$14.5 was again significant at 0.19\%$\pm$0.10\%. Not considering the first epoch, the dominant axis of \ion{Si}{{\sc II}}$\lambda6355$ was roughly constant staying close the continuum until both rotated in opposite directions on day +14.5. Detailed radiation-hydrodynamical simulations produce a very steep density slope in the outermost ejecta so that the low first-epoch $p^{\rm cont} \approx 0.2$\% nevertheless suggests a separate structure with an axis ratio $\sim$2 in the outer carbon-rich (3.5--4)$\times10^{-3}$M$_\odot$.  Large-amplitude fluctuations in the polarization profiles and a flocculent appearance of the polar diagram for the \ion{Ca}{\sc II}\, near-infrared triplet (NIR3) may be related by a common origin. The temporal evolution of the polarization spectra agrees with an off-center delayed detonation. The late-time increase in polarization and the possible change in position angle are also consistent with an aspherical $^{56}$Ni core. The $p^{\rm cont}$ and the absorptions due to \ion{Si}{{\sc II}} $\lambda6355$ and \ion{Ca}{\sc II}\,NIR3 form in the same region of the extended photosphere, with an interplay between line occultation and thermalisation producing $p$.  Small-scale polarization features may be due to small-scale structures, but many could be related to atomic patterns of the quasi-continuum; they hardly have an equivalent in the total-flux spectra. We compare SN\,2019np to other SNe and develop future objectives and strategies for SN\,Ia spectropolarimetry.

\end{abstract}


%%%%%%%%%%%%%%%%%%%%%%%%%%%%%%%%%%%%%%%%%%%%%%%%%%

%%%%%%%%%%%%%%%%% BODY OF PAPER %%%%%%%%%%%%%%%%%%

\section{Introduction} \label{sec_intro}

Various models of Type Ia supernova (SN) explosions predict photometric and spectroscopic evolution that reproduce observations adequately but not uniquely \citep{2017hsn..book.....A}, so it is difficult to judge models merely by their power in matching light curves and \textcolor{black}{total-}flux spectra.  However, they predict different explosion geometries of the progenitor white dwarf (WD), which can be diagnosed with polarimetry \citep{2021ApJ...922..186H}.  Polarized optical flux from supernovae (SNe) can be caused by departures from spherical symmetry of the global ejecta structure or by chemical ``clumps'' with different line opacities that block portions of the photosphere \citep{Wang_wheeler_2008,2017hsn..book.1017P}.  Both schemes can be understood as an incomplete cancellation of the electric vectors integrated over the photosphere as seen by the observer.  Optical polarimetry probes the geometric properties of the SN explosion and the structure of the SN ejecta, without spatially resolving the source.  A wavelength-independent continuum polarization would arise from Thomson scattering of free electrons with a globally aspherical distribution.  In addition or alternatively, it may be caused by energy input that is spatially offset from the center of mass \citep{Hoeflich_etal_1995, 1999ApJ...527L..97L, Kasen_etal_2003, Hoeflich_etal_2006}. Polarized spectral features can be induced in the SN ejecta by chemically uneven blocking within the photosphere and by frequency variations of the associated line opacities in the thermalisation depth. 
%\footnote{\bf To first order, photons thermalize at $\tau=\sqrt{3 \tau_{\rm sc} (\tau_{\rm sc}+\tau_{\rm line}) }$} 

Any early polarization signal from thermonuclear explosions offers a critical test of the nature of the progenitor systems of Type Ia SNe.  For example, large deviations from global sphericity in the density distribution and chemical abundances of the ejecta are predicted for explosions triggered by the dynamical merger of a double white dwarf (WD) binary \citep{Pakmor_etal_2012, Bulla_etal_2016a}.  The resulting polarization is expected to be significant both in the continuum and across various spectral lines.  The continuum polarization can be as high as $\gtrsim0.5$--1\% at $\sim 1$ week after the explosion if observed out of the orbital plane \citep{Bulla_etal_2016a}.  By contrast, an almost spherical density distribution and a moderate degree of chemical inhomogeneity are predicted by delayed-detonation models \citep{Hoeflich_etal_2006, Pakmor_etal_2012, Pakmor_etal_2013, Moll_etal_2014, Raskin_etal_2014}.   A continuum polarization near zero as well as modest ($\lesssim1\%$) signals across major spectral features were also predicted by specific multidimensional models for both a selected delayed-detonation and a sub-Chandrasekhar-mass (M$_{\rm Ch}$) model \citep{Bulla_etal_2016b}.

Polarized spectral lines indicate geometric deviations from spherical symmetry of the associated elements.   Chemical inhomogeneities are imprinted by the propagation of the burning front.  Delayed-detonation models predict an initial subsonic deflagration resulting in turbulence and gravitational compression.  As the burning front travels outward, the flame transforms into a supersonic detonation because of Rayleigh-Taylor instability at the interface between unburned and burned material \citep{1991A&A...245..114K}.  Layers of intermediate-mass elements (IMEs; i.e., from Si to Ca) are then produced at the front of the detonation wave.  At any given epoch, the polarization spectrum samples the geometric information of the ejecta that intersect the photosphere.  As the ejecta expand over time, the electron density decreases and the photosphere recedes into deeper layers of the ejecta in mass and velocity.  Multi-epoch spectropolarimetry tomographically maps out the distribution of various elements.

More recent early\textcolor{black}{-time} observations have also found low continuum polarization in other normal Type Ia SNe\, namely SN\,2018gv (day $-$13.6; \citealp{Yang_etal_2020}) and SN\,2019ein (day $-$10.9; \citealp{Patra_etal_2022}).  SN\,2019ein displayed one of the highest expansion velocities at early phases as inferred from the absorption minimum of the \ion{Si}{{\sc II}} $\lambda6355$ line ($\sim 24,000$\,km\,s$^{-1}$ at 14 days before photometric $B$-band maximum; \citealp{Pellegrino_etal_2020}).  The low continuum polarization on day $-$10.9 indicates a low degree of asphericity at this phase, strengthening the existing evidence that the explosions of Type Ia SNe maintain a high degree of sphericity from their early phases.  The spectropolarimetry of SN\,2018gv on day $-$13.6 was the earliest such measurement at its time for any Type Ia SN.  The 0.2\%$\pm$0.13\% continuum polarization five days after the explosion (based on phase estimates from the early light curve) suggests that the photosphere was moderately aspherical with an axis ratio of 1.1--1.3. 
\footnote{Throughout the paper, the term equatorial plane is defined by planes \textcolor{black}{through the center, $\vec{n} \vec{x}=0, $ $\vec{x}$ spanning the plane with the symmetry axis of a rotationally symmetric ellipsoid as the orthogonal vector $\vec{n}$, or $\vec{n}$ being} a line through the center of the WD and the location of an off-center energy source \citep{1995ApJ...443...89H}.  The two so defined planes may be different (Section~\ref{sec_model}).
}.
However, even at this early phase, the geometry of the outermost $\sim 10^{-3}$ to $\sim 10^{-2}$ M$_{\rm WD}$ of SN\,2018gv still remained observationally unconstrained.  The polarization is also sensitive to the rapidly-changing density structure in the outer layers, which intersect the photosphere in the first few days \citep{Hoeflich_etal_2017}.

SN\,2019np was discovered at 2019-01-09 15:58 \textcolor{black}{(UT dates are used throughout this paper)} with a 0.5\,m telescope at a clear-band magnitude of 17.8 \citep{Itagaki_2019}.  Rapid spectroscopic follow-up \textcolor{black}{observations were} carried out as early as $\sim 1$ day after the discovery \citep{Kilpatrick_etal_2019, Burke_etal_2019, Wu_etal_2019}.  Spectral cross-correlations with the ``Supernova Identification'' (SNID; \citealp{Blondin_etal_2007}) and the ``Superfit'' \citep{Howell_etal_2005} codes suggest that SN\,2019np is a Type Ia SN discovered $\sim 2$ weeks before maximum light.  From the photometry by \citet{Burke_etal_2022}, we derived that SN\,2019np reached its peak $B$-band magnitude at MJD 58509.72$\pm$0.06$\pm$0.51 (see Appendix.~\ref{sec_earlylc}), where the two uncertainties represent the statistical and the systematic error, respectively. This estimate is consistent with the respective values of 58510.2$\pm$0.8 and 58509.64$\pm$0.06 reported by \citet{Sai_etal_2022} and \citet{Burke_etal_2022}.  All phases used throughout the present paper are given relative to the $B$-band maximum light at MJD 58509.72 (2019-01-26.72).  A comprehensive study of the SN by \citet{Sai_etal_2022} concluded that its photometric and spectroscopic properties were similar to those of other normal Type Ia SNe.

\citet{Sai_etal_2022} detected a $\lesssim$5\% excess in the early bolometric flux evolution of SN\,2019np compared to radiative diffusion models \citep{Arnett_1982, Chatzopoulos_etal_2012}, hinting at additional energy input compared to the radioactive decay of a Ni core.  They suggested that the blue and relatively fast-rising early light curves of SN\,2019np are best fitted with the mixing of $^{56}$Ni from the inner to the outer layers of the SN ejecta \citep{Piro_etal_2016}.  The rise time of SN\,2019np is not compatible with models that predict an early interaction between the SN ejecta and any ambient circumstellar matter (CSM) or a companion star \citep{Kasen_2010}.  Moreover, the colour evolution of SN\,2019np is inconsistent with that predicted for a progenitor WD below $M_{\rm Ch}$ and surrounded by a thin helium shell \textcolor{black}{as discussed in Sections \ref{sec_other_sne} \& \ref{sec_opportunities}} .  In this ``double-detonation'' or ``He-shell detonation'' picture, an initial detonation is triggered in the surface He shell, sending a shock wave to the inner region of the C/O WD.  The shock generates compression heat and subsequently triggers the second detonation that ignites the WD \citep{Woosley_etal_1980, Nomoto_1982_p1, Nomoto_1982_p2, Livne_1990, Woosley_Weaver_1994, Hoeflich_Khokhlov_1996, Kromer_etal_2010}.  \citet{Burke_etal_2022} also suggested an excess in the early flux evolution of SN\,2019np, which may have been too weak to have been caused by an interaction between the ejecta and a companion. Interaction with any CSM is an additional possibility. 

This study presents five epochs of optical spectropolarimetry of SN\,2019np from $t \approx -14.5$ to +14.5 days and interpretations based on detailed radiation-hydrodynamic simulations.  The paper is organised as follows.  In Section~\ref{sec_obs} we outline the spectropolarimetric observations and the data-reduction procedure.  The polarization properties of SN\,2019np are discussed in Section~\ref{Sec_spec}.  The analysis of these properties with hydrodynamic models is carried out in Section~\ref{sec_model}.  We summarise our conclusions in Section~\ref{sec_conclusions}, and develop a comprehensive appraisal of the potential of spectropolarimetry for the understanding of Type Ia SNe in Section~\ref{sec_summary}.

\section{Spectropolarimetry of SN\,2019\lowercase{np}}~\label{sec_obs}

Spectropolarimetry of SN\,2019np was conducted with the FOcal Reducer and low dispersion Spectrograph 2 (FORS2; \citealp{Appenzeller_etal_1998}) on Unit Telescope~1 (UT1, Antu) of the ESO Very Large Telescope (VLT).  The Polarimetric Multi-Object Spectroscopy (PMOS) mode was used for all science observations.  A complete set of spectropolarimetry consists of four exposures at retarder-plate angles of 0, 22.5, 45, and 67.5 degrees.  The 300V grism and a 1$\arcsec$-wide slit were selected for all observations.  The order-sorting filter GG435 was in place, which has a cut-on at $\sim$4350 \AA\ to prevent shorter-wavelength second-order contamination.  This configuration provides a spectral resolving power of $R \approx 440$ at a central wavelength of 5849\,\AA, corresponding to a resolution-element size of $\sim 13$\,\AA\ (or $\sim670$\,km\,s$^{-1}$) according to the VLT FORS2 user manual \citep{Anderson_etal_2018}. 

Observations were obtained at five epochs: (in the format day/UT) $-$14.5/2019-01-12, $-$11.4/2019-01-15, $-$6.4/2019-01-20, $+$0.5/2019-01-27, and $+$14.5/2019-02-10.  At the first epoch, the total $4 \times 1100$\,s integration time was split into two sets of exposures to reduce the impact of cosmic rays.  The two loops were carried out at relatively large and different airmasses, from 1.84 to 1.73 and from 1.73 to 1.70.  We conducted a consistency check of the two measurement sets and found that the Stokes parameters derived for the two loops agree within their 1$\sigma$ uncertainties over the entire wavelength range after rebinning the data to larger resolution elements (e.g., 30\,\AA\ and 40\,\AA\ bin sizes).  We thus combined the two datasets by taking the mean value of the spectra obtained at each retarder-plate angle.  Relative-flux calibration was based on the flux standard star HD\,93621 observed at a half-wave plate angle 0 degrees near epoch 3.  The airmass of the flux standard was chosen to be comparable to that of the spectropolarimetry of SN\,2019np.  A log of the VLT spectropolarimety is presented in Table~\ref{Table_log_specpol}. 

After bias and flat-field corrections, the ordinary (o) and extraordinary (e) beams in each two-dimensional spectral image were extracted following standard routines within IRAF\footnote{{IRAF} is distributed by the National Optical Astronomy Observatories, which are operated by the Association of Universities for Research in Astronomy, Inc., under cooperative agreement with the National Science Foundation.} \citep{Tody_1986, Tody_1993}.  A typical \textcolor{black}{root-mean-square (RMS)} accuracy of $\sim 0.20$\,\AA\ was achieved in the wavelength calibrations.  Stokes parameters were then derived using our own routines based on the prescriptions in \citet{Patat_etal_2006_polerr} and \citet{Maund_etal_2007_05bf}, which also correct the bias due to the non-negativity of the polarization degree.  The observed polarization degree and position angle ($p_{\rm obs}$, {\it PA}$_{\rm obs}$) and the true values after bias correction ($p$, {\it PA}) can be written as
\begin{equation}
\begin{aligned}
p_{\rm obs} = \sqrt{Q^2 + U^2}, \ 
%{\bf  p = p_{\rm obs} - \frac{\sigma_{p}^2}{p_{\rm obs} h(p_{\rm obs} - \sigma_p)}} \\
{p = \bigg{(} p_{\rm obs} - \frac{\sigma_{p}^2}{p_{\rm obs}}\bigg{)} \times h(p_{\rm obs} - \sigma_p);} \\
%p = (p_{\rm obs} - \sigma_{p}^2 / p_{\rm obs}) \times h(p_{\rm obs} - \sigma_p); \\
PA_{\rm obs} = \frac{1}{2} {\rm arctan} \bigg{(} \frac{U}{Q} \bigg{)}, 
\ {\rm \ and \ } PA = PA_{\rm obs}. 
\end{aligned}
\label{Eqn_stokes0}
\end{equation}
Here, $Q$ and $U$ are the intensity ($I$)-normalised Stokes parameters.  The correction for the polarization bias is based on equations in \citet{Simmons_etal_1985} and \citet{Wang_etal_1997}, where $\sigma_{p}$ and $h$ denote the $1\sigma$ uncertainty in $p_{\rm obs}$ and the Heaviside step function, respectively.  {The brackets are properly set as in \citet{Cikota_etal_2019}.}  

A $\lesssim 0.1$\% instrumental polarization was also corrected following the procedure discussed by \citet{Cikota_etal_2017_fors2}.  More details of the reduction of FORS2 spectropolarimetry can be found in the FORS2 Spectropolarimetry Cookbook and Reflex Tutorial\footnote{\url{ftp://ftp.eso.org/pub/dfs/pipelines/instruments/fors/fors-pmos-reflex-tutorial-1.3.pdf}}, as well as in \citet{Cikota_etal_2017_fors2} and \citet{Yang_etal_2020}.

\section{Polarimetric Properties of SN\,2019\lowercase{np}}~\label{sec_spec_pol}
\label{Sec_spec}

\begin{figure}
\includegraphics[width=1.0\linewidth]{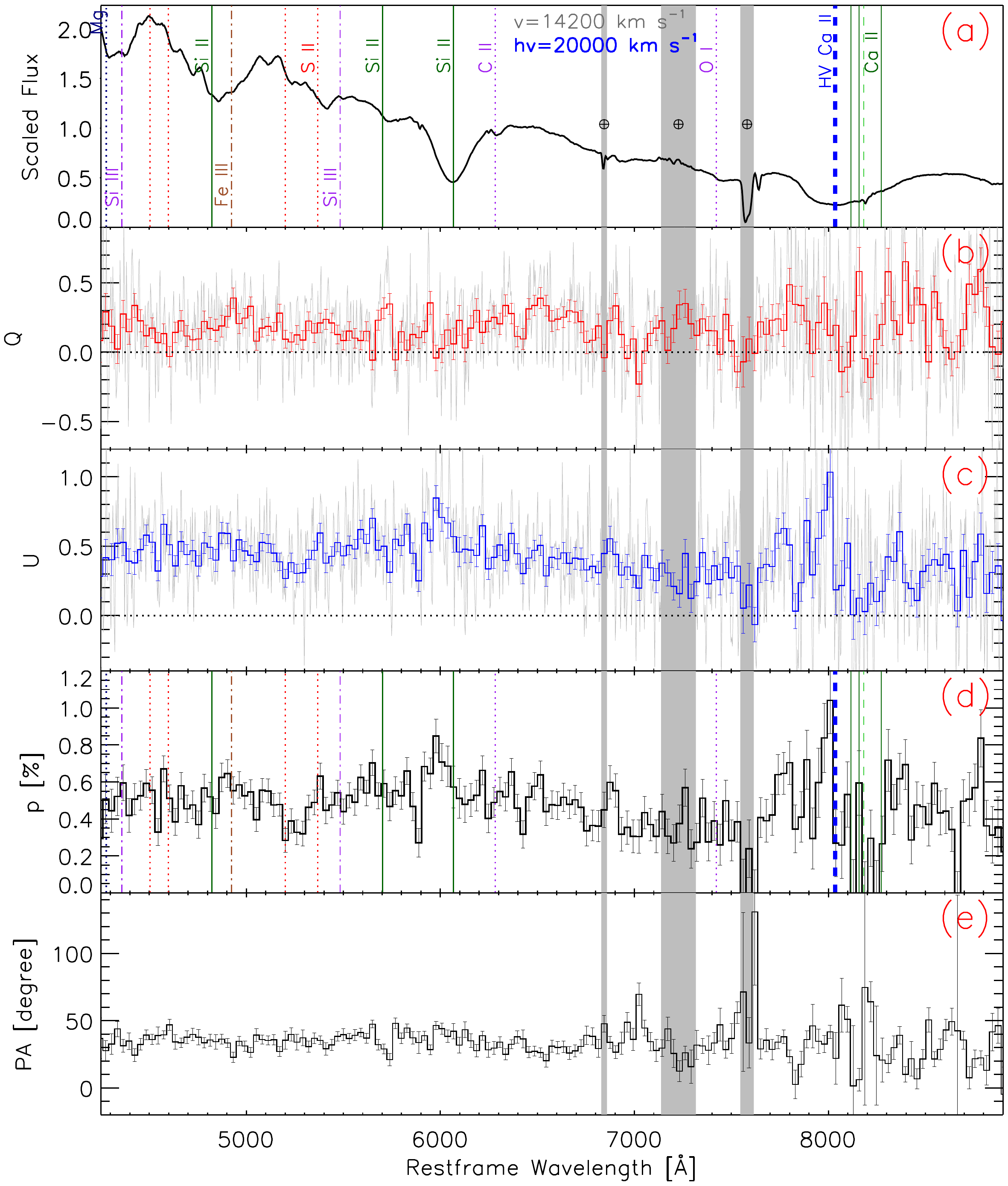}
\caption{Spectropolarimetry of SN\,2019np on day $-$14.5 (epoch 1) relative to $B$-band maximum light on MJD 58509.7.  The five panels (from top to bottom) display (a) the arbitrarily scaled flux spectrum with major spectral features identified and the high-velocity component of \ion{Ca}{{\sc II}}\,NIR3 labeled ``hv''; (b,c) the normalised Stokes parameters $Q$ and $U$, respectively; (d) the polarization spectrum ($p$); and (e) the polarization position angle.  Panels (b)--(e) represent the polarimetry before ISP correction.  The grey lines in panels (b) and (c) show the data with their original sampling while the heavy lines in panels (b)--(e) use 30\,\AA\ bins for clarity.  The grey-shaded vertical bands identify regions of telluric contamination.
\label{Fig_iqu_ep1}
}
\end{figure}

\begin{figure}
   \begin{minipage}[t]{0.48\textwidth}
     \centering
     \includegraphics[width=1.02\linewidth]{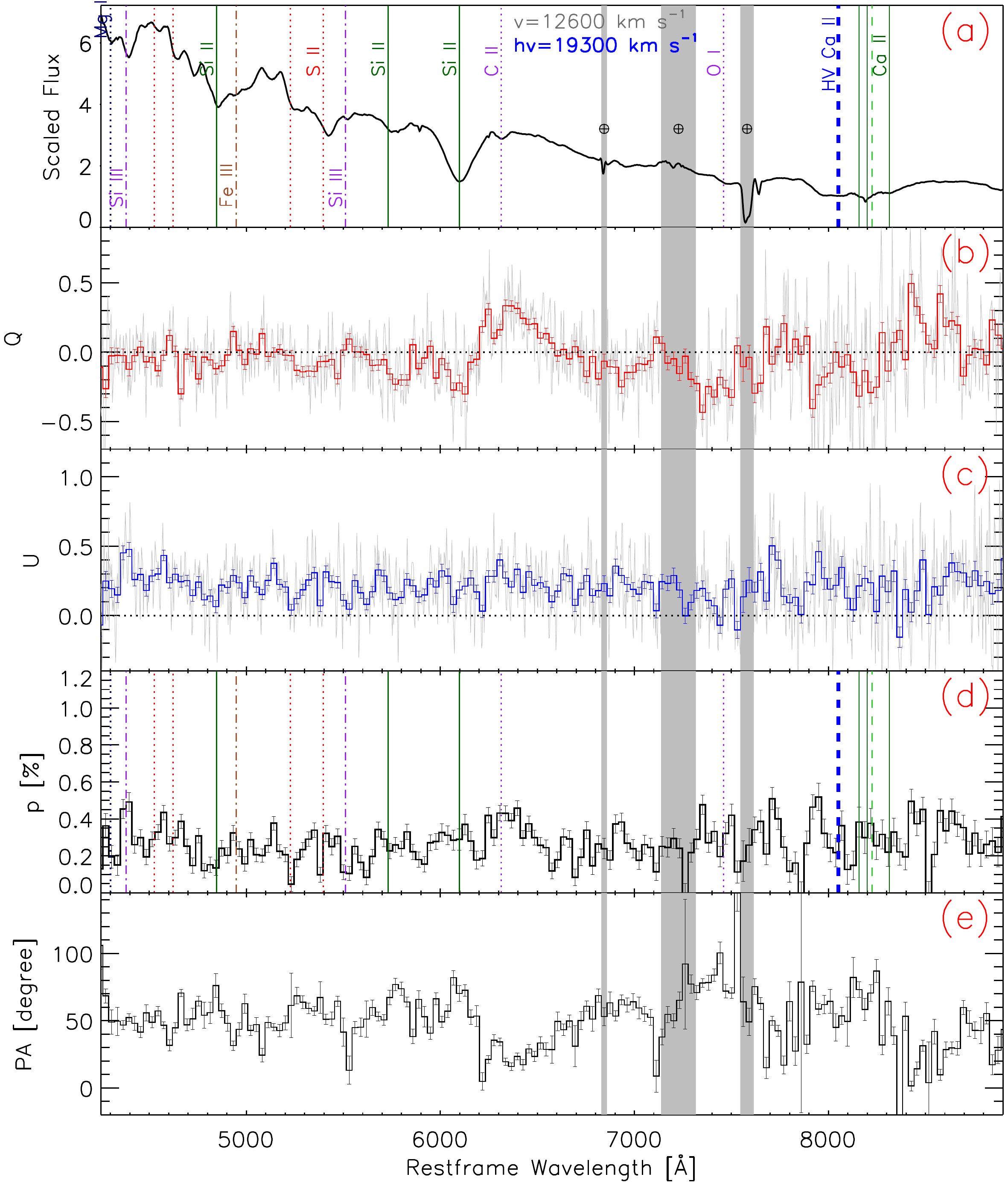}
     \caption{Same as Figure~\ref{Fig_iqu_ep1}, but for day $-$11.4 (epoch 2).}\label{Fig_iqu_ep2}
   \end{minipage}\hfill
   \begin{minipage}[t]{0.48\textwidth}
     \centering
     \includegraphics[width=1.02\linewidth]{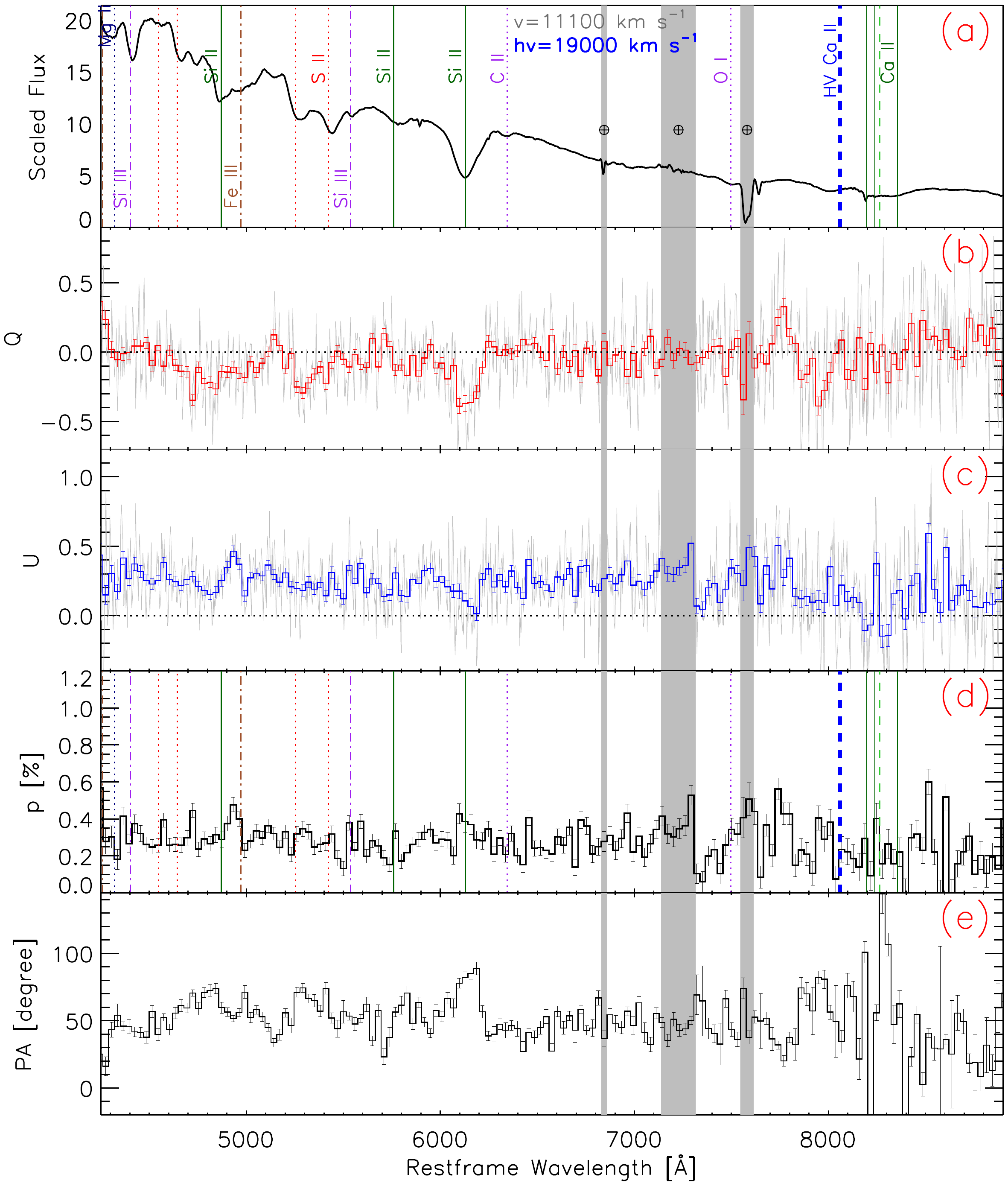}
     \caption{Same as Figure~\ref{Fig_iqu_ep1}, but for day $-$6.4 (epoch 3).}\label{Fig_iqu_ep3}
   \end{minipage}
\end{figure}

\begin{figure}
   \begin{minipage}[t]{0.48\textwidth}
     \centering
     \includegraphics[width=1.02\linewidth]{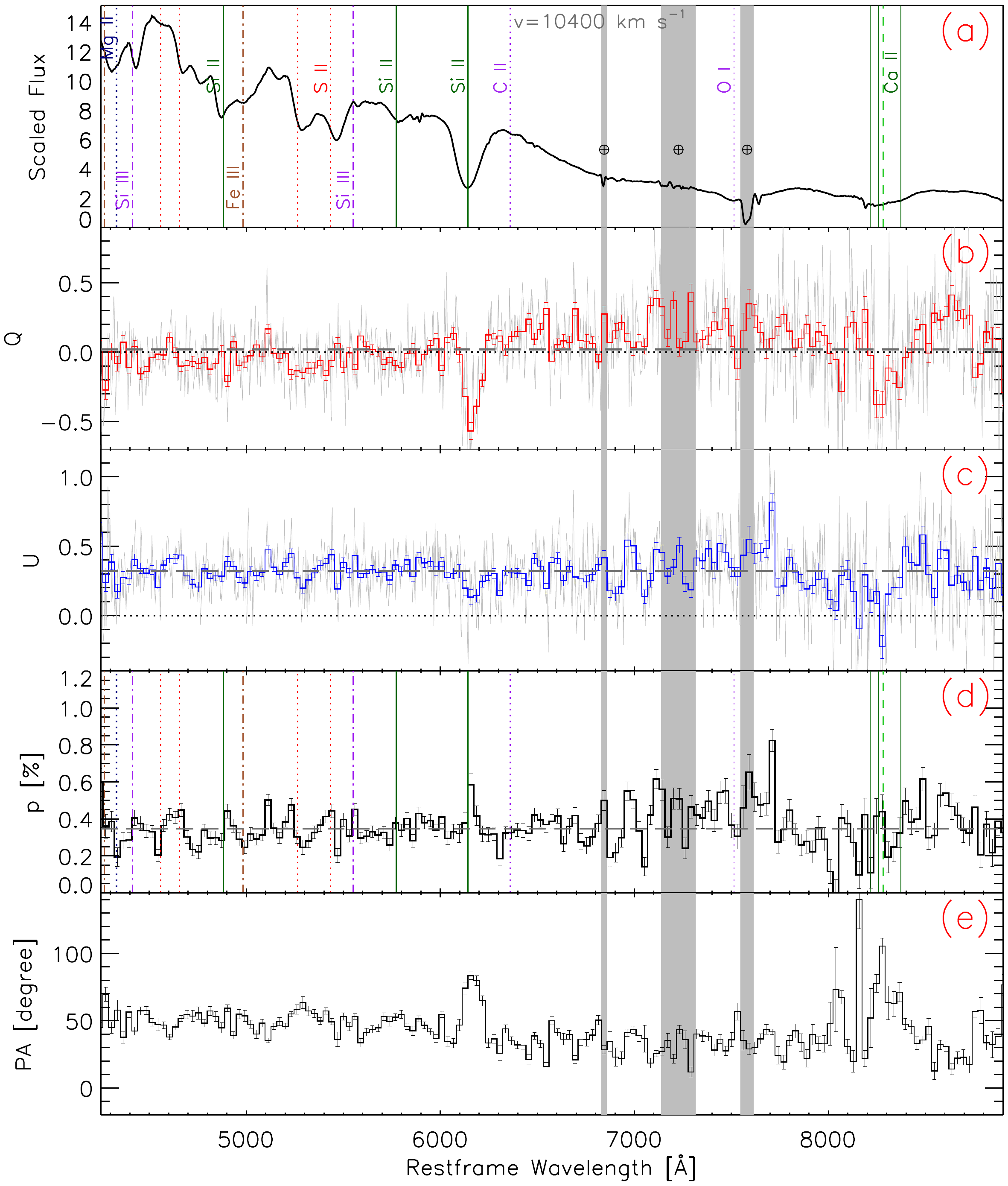}
     \caption{Same as Figure~\ref{Fig_iqu_ep1}, but for day $+$0.5 (epoch 4). The estimated ISP level is shown by grey-dashed lines in panels (b)--(d).}\label{Fig_iqu_ep4}
   \end{minipage}\hfill
   \begin{minipage}[t]{0.48\textwidth}
     \centering
     \includegraphics[width=1.02\linewidth]{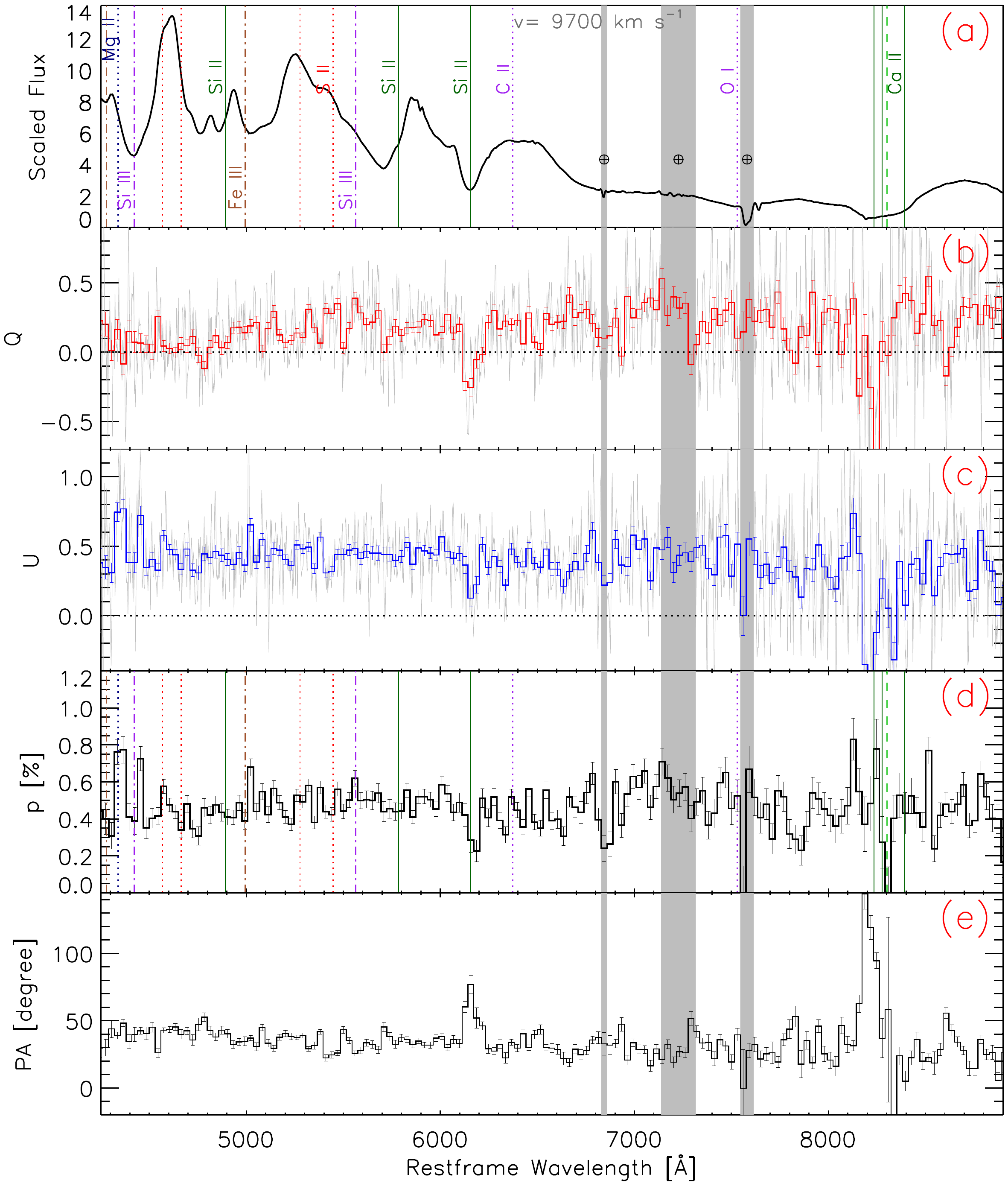}
     \caption{Same as Figure~\ref{Fig_iqu_ep1}, but for day $+$14.5 (epoch 5).}\label{Fig_iqu_ep5}
   \end{minipage}
\end{figure}

The spectropolarimetry of SN\,2019np obtained on days $-$14.5, $-$11.4, $-$6.4, $+$0.5, and $+$14.5 is presented in Figs.~\ref{Fig_iqu_ep1}--\ref{Fig_iqu_ep5}, respectively, where the data are not corrected for interstellar polarization (ISP).  Polarization spectra are shown together with the associated scaled \textcolor{black}{total-}flux spectra \textcolor{black}{(hereafter referred to as simply ``flux spectra'')}.  Both have been transformed to the rest frame. 

\begin{figure}
\includegraphics[width=1.0\linewidth]{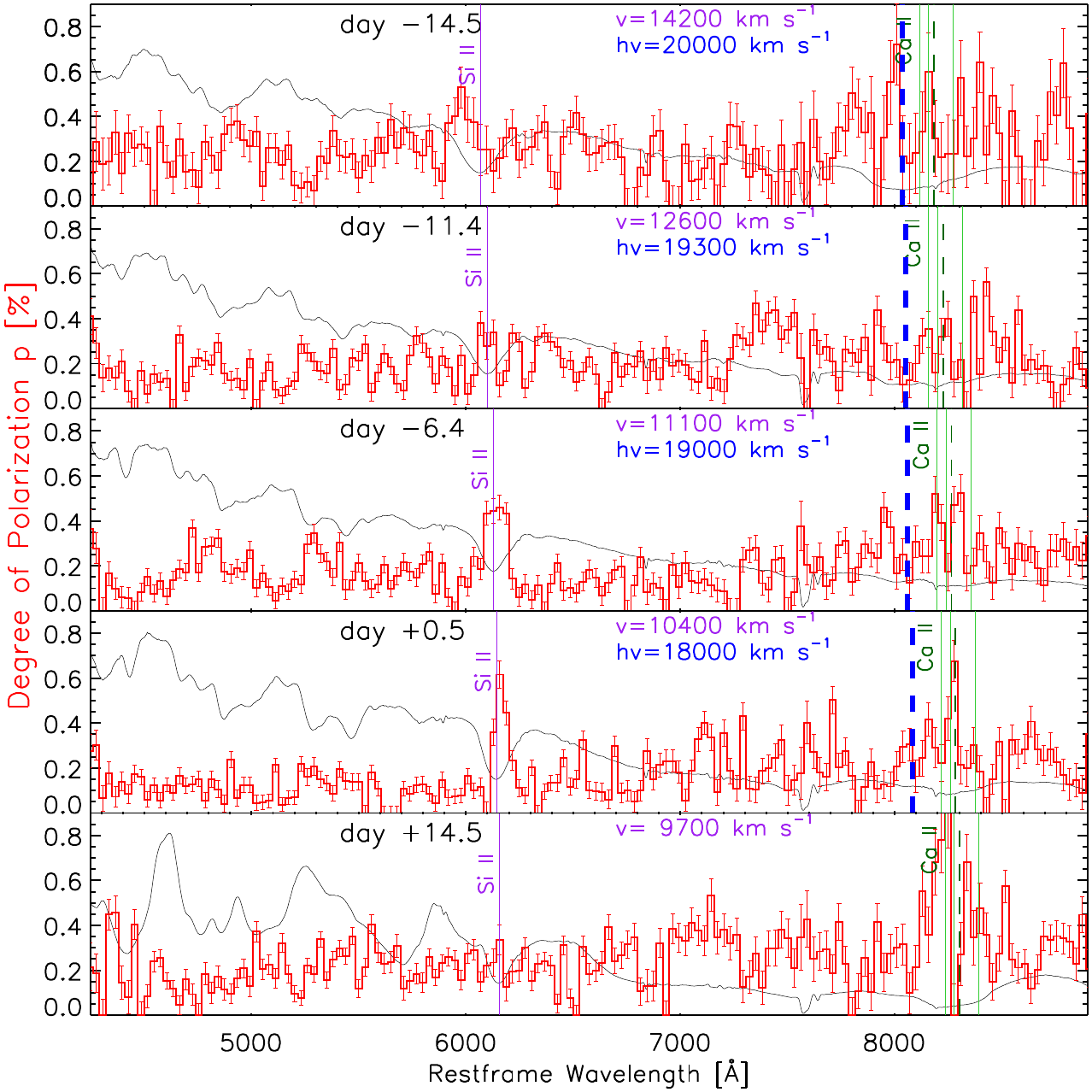}
\caption{Intrinsic polarization of SN\,2019np from days $-$14.5 to $+$14.5 as labeled from top to bottom in the subpanels. 
{ For each epoch, the degree of polarization is calculated based on ISP-subtracted Stokes $Q$ and $U$, bias-corrected following Equation~\ref{Eqn_stokes0}, and presented} (red histograms) with 30\,\AA\ binning, together with the arbitrarily scaled flux spectrum (black lines).  \ion{Si}{{\sc II}} $\lambda$6355 and the photospheric component of the \ion{Ca}{{\sc II}}\,NIR3 features are marked, and their velocities ($v$) are also given.  The high-velocity component of the \ion{Ca}{{\sc II}}\,NIR3 feature is labeled ``hv.''
\label{Fig_pol}
}
\end{figure}

\subsection{Interstellar Polarization}~\label{sec_isp}
Removing the polarization imposed by interstellar dust grains in either the Milky Way or the host galaxy or both is essential for revealing the intrinsic polarization of SNe.  This ISP is due to dichroic extinction by nonspherical dust grains aligned by the interstellar magnetic field. \textcolor{black}{Therefore, the entire observed wavelength range of the spectrum is used to determine the overall level of the ISM polarization. {\textcolor{black}{As will be shown in Section \ref{Sec_pol_spec}, both the overall level and the continuum polarization  in a narrow wavelength range plus the spectral features are consistent in our analysis. This provides an argument that the procedure to find the ISM polarization does not suppress an overall net-polarization mimicking overall sphericity.}}} The intrinsic continuum polarization of Type Ia SNe around their peak luminosity is very low ($\lesssim 0.2$\%; see, e.g., \citealp{Wang_wheeler_2008,2017hsn..book.1017P,Yang_etal_2020}). 
Therefore, we used the spectrum of SN\,2019np from day $+$0.5 as an unpolarized standard.  We fitted the Stokes $Q$, $U$ parameters and the observed degree of polarization, $p$, using Serkowski's wavelength-dependent law \citep{Serkowski_etal_1975} as well as a mere constant.  In the given low-ISP regime, we found that Serkowski's law failed to yield a satisfactory fit, and the ISP can be characterised by the latter approach, which requires computing the error-weighted mean values of $Q$ and $U$ over suitably selected spectral regions.  Using the wavelength range 4400--8900\,\AA\, but excluding the telluric features and the strongly polarized \ion{Si}{{\sc II}} $\lambda$6355 line and the \ion{Ca}{{\sc II}}\, \textcolor{black}{near-infrared (NIR) triplet} (8500.36\,\AA, 8544.44\,\AA, and 8664.52\,\AA, with a central wavelength of $\lambda_{0} \approx 8570$, denoted as \ion{Ca}{{\sc II}}\,NIR3 hereafter) due to the SN, we estimate the ISP as ($Q_{\rm ISP}$, $U_{\rm ISP}$) = (0.019$\pm$0.121\%, 0.322$\pm$0.072\%), and $p_{\rm ISP} = 0.343\pm0.075$\%.  These values are well consistent with the ISP derived over the wavelength ranges which are considered to be depolarized due to blanketing by numerous iron absorption lines (see, e.g., \citealp{Howell_etal_2001, 2006NewAR..50..470H, Patat_etal_2008, Maund_etal_2013, Patat_etal_2015, Yang_etal_2020}).  Adopting the Galactic and the host-galaxy reddening of SN\,2019np of $E(B-V)_{\rm Gal} = 0.018$\,mag and $E(B-V)_{\rm host} = 0.10 \pm 0.03$\,mag \citep{Sai_etal_2022}, we find the estimated ISP consistent with the empirical upper limit caused by dichroic extinction and established for dust in the Galaxy, $p_{\rm ISP} \textless 9 \% \times E(B−V)$, following \citet{Serkowski_etal_1975}.

\subsection{Intrinsic Continuum Polarization}~\label{sec_contpol_obs}

After subtracting the ISP, we determined the continuum polarization of SN\,2019np at all epochs from the Stokes parameters over the wavelength range 
6400--7000\,\AA, which is considered to be free of significant polarized spectral features \citep{Patat_etal_2009}.
The error-weighted mean Stokes parameters within this region are given in Table~\ref{Table_pol}.  The uncertainty was estimated by adding the statistical errors and the standard deviation computed from the 30\,\AA-binned spectra within the chosen wavelength range in quadrature.  The continuum polarization within this wavelength interval is consistent with that computed over the entire observed wavelength range after exclusion of the broad, polarized \ion{Si}{{\sc II}} $\lambda$6355 and \ion{Ca}{{\sc II}}\,NIR3 lines. 

The intrinsic continuum polarization of SN\,2019np on day $-$14.5 was 0.21\%$\pm$0.09\%.  After only three days, it had dropped to $\sim 0$ by day $-$11.4 and remained low until the SN reached its peak luminosity.  By day $+$14.5, the continuum polarization had increased to 0.19\%$\pm$0.10\%.\footnote{\textcolor{black}{Note that the time variation in $p^{\rm cont}$ is at a 2 $\sigma$ level. However, the significance of the variation is strongly supported by the change in the  dominant axes in the $Q$--$U$ plane. Moreover, the estimate of the uncertainty in $p^{\rm cont}$ includes real spectral variations in $p$ caused by spectral features (see Sections \ref{sec_contpol_obs}, \ref{sec_quplane}, and \ref{Sec_pol_spec}}).}   From the power-law fit of the earliest light curves of SN\,2019np, we place the time of first light at $t_{0} = -17.92 \pm 0.06$ day, where 0.06 day is only the statistical error (see Appendix~\ref{sec_earlylc}).  An additional systematic error of $\pm$0.51 day results from the determination of the time of the peak luminosity.  The times of the five epochs of VLT spectropolarimetry relative to this time of the SN explosion are 3.5, 6.5, 11.5, 18.5, and 32.4 days, respectively.  The first epoch is the earliest such measurement for any Type Ia SN to date.

{
\subsection{The Dominant Axes in the {\boldmath \textit{Q--U}} Plane}~\label{sec_quplane}
}

For each epoch of our ISP-corrected spectropolarimetry, we examine in the Stokes $Q$--$U$ plane the axial symmetry of the ejecta of SN\,2019np as they enter the extended photosphere.  We do this separately for suitable wavelength ranges covering the continuum and the \ion{Si}{{\sc II}} $\lambda$6355 and \ion{Ca}{{\sc II}}\,NIR3 lines.  This method was introduced by \citet{Wang_etal_2001}; a different graphical rendering of the same data will be discussed in Section~\ref{sec_line_pol}.  Purely axially symmetric ejecta imprint a linear structure on the $Q$--$U$ plane, since the orientation of the structure is defined by a single polarization position angle, while varying scattering and polarization efficiencies lead to deviations from a straight line.  By projecting the data onto the best-fitting axis and measuring the scatter about this so-called dominant axis,
\begin{equation}
U = \alpha + \beta Q\,, 
\label{Eqn_daxis}
\end{equation}
one may characterise the degree of axial symmetry of the SN ejecta \citep{Wang_etal_2003_01el, Maund_etal_2010_05hk}. 

\begin{figure*}
\includegraphics[width=1.0\linewidth]{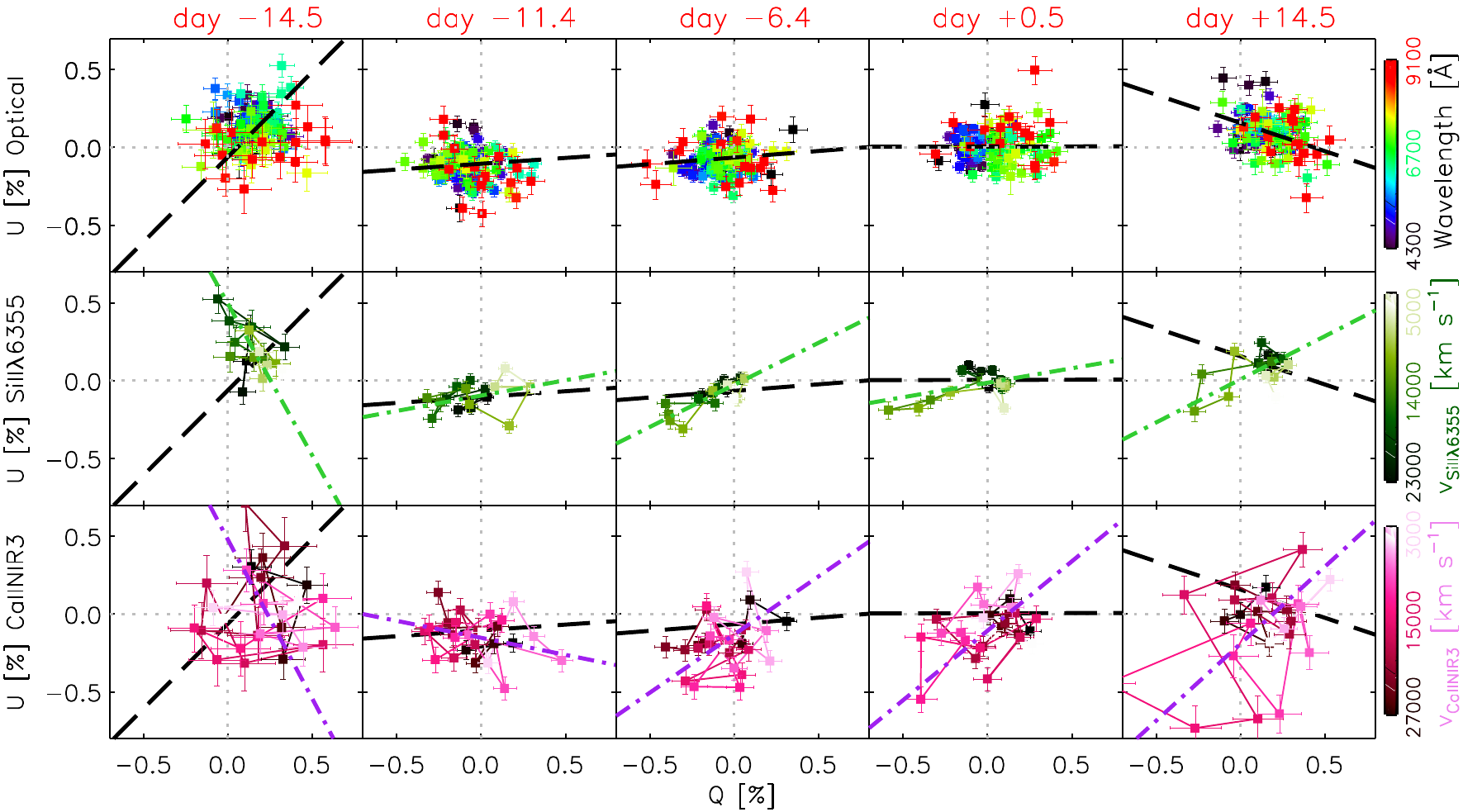}
	\caption{Intrinsic polarization of SN\,2019np displayed on the Stokes $Q$--$U$ plane.  The top row shows the data over the wavelength range 4250--9100\,\AA.  The wavelength of each 30\,\AA\ bin is indicated by the colour bar on the right side.  The middle and the bottom rows display the polarization for the \ion{Si}{{\sc II}} $\lambda$6355 and \ion{Ca}{{\sc II}}\,NIR3 features over the velocity ranges of 24,000--4000\,km\,s$^{-1}$ and 28,000--2000\,km\,s$^{-1}$, respectively.  The velocities are also indicated by the corresponding colour bars.  The epochs of the observations are labeled with their phases at the top of each column.  In each panel, the black long-dashed line shows the dominant axis calculated over the wavelength range 4250--9100\,\AA\ with the \ion{Si}{{\sc II}} $\lambda$6355 and \ion{Ca}{{\sc II}}\,NIR3 features excluded (the values of the fitting parameters $\alpha$ and $\beta$ in eq. \ref{Eqn_daxis} are listed in Table~\ref{Table_pol}).  In the middle and the bottom rows, the green and purple dot-dashed lines in each subpanel represent linear fits to the displayed data points that cover the \ion{Si}{{\sc II}} $\lambda$6355 and the \ion{Ca}{{\sc II}}\,NIR3 features, respectively.
\label{Fig_qu}
}

\includegraphics[width=1.0\linewidth]{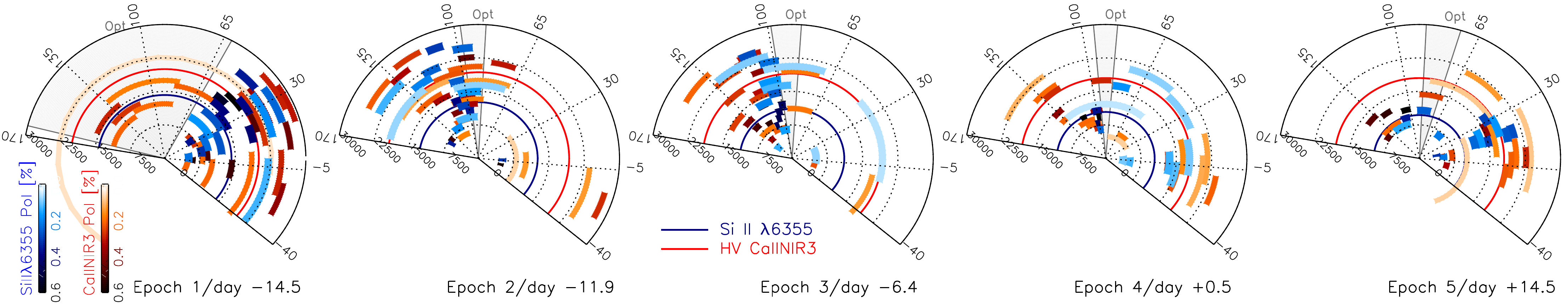}
\caption{Polar plots of the polarization of SN\,2019np across the \ion{Si}{{\sc II}} $\lambda$6355 and \ion{Ca}{{\sc II}}\,NIR3 lines at all five epochs.  In each panel, the radial distance and the angle represent the expansion velocity and the polarization position angle, respectively.  The center of each fan-shaped bin gives the average position angle calculated over the velocity range covered by the radial extent of the bin.  The angular widths of the fan-shaped bins represent the 1$\sigma$ uncertainty on the position angle rather than the underlying physical dimensions.  The velocity is labeled in km\,s$^{-1}$, and the celestial position angles are measured in degrees from North to East.  The blue and the orange colour bars indicate the ISP-corrected polarization degree across the \ion{Si}{{\sc II}} $\lambda$6355 and \ion{Ca}{{\sc II}}\,NIR3 profiles, respectively.  The data have been rebinned to 30\,\AA\ for better visualisation.  The direction of the dominant axis is shown by grey-shaded regions with their angular width representing the 1$\sigma$ uncertainty (i.e., $\theta_{d}^{*}$ in Table~\ref{Table_pol}).  The blue and red semicircles mark the estimated photospheric velocity and the high-velocity component as measured from the \textcolor{black}{absorption minima of} the \ion{Si}{{\sc II}} $\lambda$6355 and \ion{Ca}{{\sc II}}\,NIR3 lines, respectively. 
\label{Fig_polar}
}
\end{figure*}

Figure~\ref{Fig_qu} displays the ISP-corrected Stokes parameters on the $Q$--$U$ plane between days $-$14.5 and $+$14.5.  The dominant axis of {SN\,2019np as determined from its polarization projected on the $Q$--$U$ plane} was derived by performing an error-weighted linear least-squares fit to the entire observed wavelength range ($4350 \leq \lambda \leq 9100$\,\AA) with the prominent and polarized \ion{Si}{{\sc II}} $\lambda$6355 and \ion{Ca}{{\sc II}}\,NIR3 features excluded.  Data points covering the \ion{Si}{{\sc II}} $\lambda$6355 and \ion{Ca}{{\sc II}}\,NIR3 profiles were omitted in the top row, where the dominant axis appears as the black long-dashed line. 

To examine the difference between the fits including and excluding the \ion{Si}{{\sc II}} $\lambda$6355 and \ion{Ca}{{\sc II}}\,NIR3 lines, we list the dominant axis and the corresponding position angles for both cases in Table~\ref{Table_pol}.  The dominant axes of SN\,2019np fitted for both cases are consistent with each other within their 1$\sigma$ uncertainties except for epochs 3 and 4, when SN\,2019np reached its peak luminosity and the discrepancy between the two fits amounts to $\sim 2\sigma$.  We consider the fits with both broad and polarized individual features excluded a more reasonable characterisation of the orientation of the 
SN ejecta since these \textcolor{black}{Si and Ca} features generally exhibit significant deviations from the rest of the wavelength range \citep{Leonard_etal_2005}.

In the middle and bottom rows of Figure~\ref{Fig_qu}, the directions of the symmetry axes of the \ion{Si}{{\sc II}} $\lambda$6355 and \ion{Ca}{{\sc II}}\,NIR3 features are shown by the green and purple dot-dashed lines, respectively.  The fitting procedures were the same as for the continuum but over the velocity ranges from 24,000 to 0\,km\,s$^{-1}$ for \ion{Si}{{\sc II}} $\lambda$6355 and from 28,000 to 0\,km\,s$^{-1}$ for the \ion{Ca}{{\sc II}}\,NIR3 complex.  The derived parameters are also listed in Table~\ref{Table_pol}.  On day $-$14.5, the spectropolarimetry over the optical range is poorly represented by a dominant axis.  The
\ion{Ca}{{\sc II}}\,NIR3 feature is barely described by the linear fits.  Additionally, as shown by the $Q$--$U$ diagrams for day $-$14.5, data points across \ion{Si}{{\sc II}} $\lambda$6355 deviate from the clustering in the continuum, indicating a conspicuous polarization across the line.  However, \textcolor{black}{owing} to the relatively low signal-to-noise ratio \textcolor{black}{(SNR)} and the moderate level of polarization, it is hard to quantitatively determine whether \ion{Si}{{\sc II}} $\lambda$6355 and the ejecta of SN\,2019np determined from the optical continuum (as far as recorded by FORS2) follow different geometric configurations. 

Starting from day $-$11.4, the ejecta of SN\,2019np have developed a more discernible symmetry axis compared to day $-$14.5.  This is indicated by the significantly reduced uncertainties in the linear least-squares fits to the polarimetry on the $Q$--$U$ plane (see the $\alpha^{*}$, $\beta^{*}$, and $\theta_{d}^{*}$ values in Table~\ref{Table_pol}). The dominant axis of SN\,2019np shows little temporal evolution between days $-$11.4 and $+$0.5 and rotates by $\sim$15$^{\circ}$ from days $+$0.5 to $+$14.5.
\textcolor{black}{Polar diagrams for \ion{Ca}{\sc II}\,NIR3 appeared very flocculent, and somewhat misaligned with the dominant axes of \ion{Si}{{\sc II}} $\lambda6355$ and $p^{\rm cont}$. Qualitatively, the temporal evolution of \ion{Si}{{\sc II}} $\lambda$6355 and \ion{Ca}{{\sc II}}\,NIR3 features are similar.
This can be seen from the middle and bottom rows of Figure~\ref{Fig_qu}. 
Not considering the first epoch, the dominant axis of \ion{Si}{{\sc II}} $\lambda6355$ was \textcolor{black}{roughly} constant and stayed close to that of the continuum until both rotated in opposite directions on day +14.5.}
 Not considering day $-$14.5, we suggest that SN\,2019np belongs to the spectropolarimetric type D1 \citep{Wang_wheeler_2008}, in which a dominant axis can be determined while the scatter of the data points about the dominant axis is conspicuous.  At the earliest epoch, a dominant axis cannot be clearly identified, and the continuum polarization measurements cluster around a location offset from the origin.

Apart from \ion{Si}{\sc II} $\lambda$6355 and \ion{Ca}{\sc II}\,NIR3, there are numerous minor peaks scattered all over the polarization spectra.  Nominally, these features are significant at $\sim 2\sigma$ and occasionally at $3\sigma$.  Careful quality control of the data and our reduction procedures have not identified them as artifacts, although some of them will undoubtedly be spurious.
Most of them are volatile and, in consecutive observations, do not appear at the same location.
{\textcolor{black}{This can be expected because the spectral features form in layers with different abundances (see Section \ref{Sec_pol_spec}).}}
In our analysis in Section~\ref{sec_model}, we will refer to them as ``wiggles''. 

\begin{figure}
\includegraphics[width=1.0\linewidth]{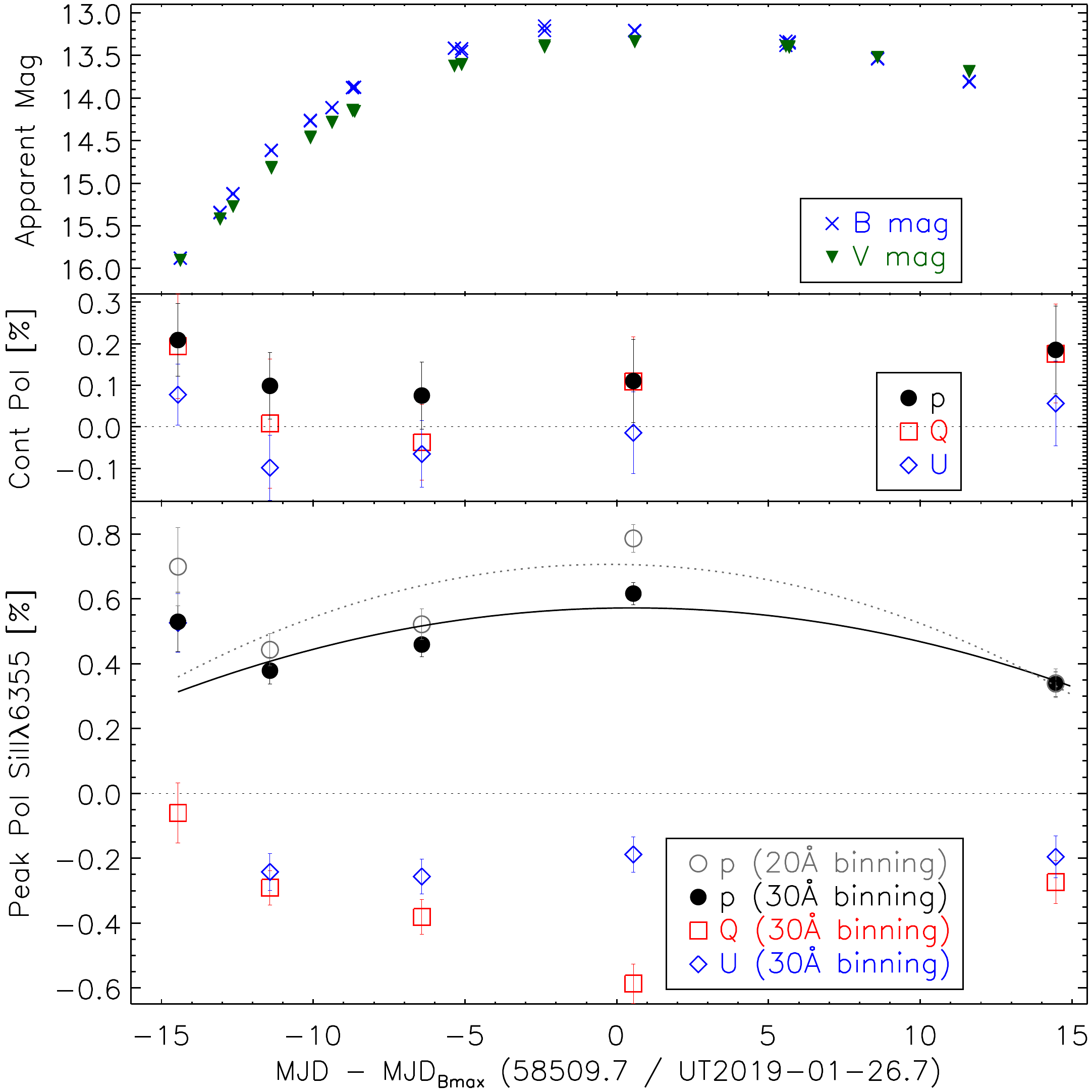}
\caption{Temporal evolution of the intrinsic polarization of SN\,2019np from days $-$14.5 to $+$14.5. The $B$ and $V$ light curves \citep[from][]{Burke_etal_2022} are displayed in the top panel.  The middle panel gives the continuum polarization calculated from the error-weighted mean values of the Stokes parameters in the range 6400--7000\,\AA.  The bottom panel presents the peak polarization measured across the \ion{Si}{{\sc II}} $\lambda$6355 line.  Values measured with both 30\,\AA\ and 20\,\AA\ bin sizes are plotted as labeled.  Second-order polynomial fits to either bin size are indicated by the solid black and dotted grey curves, respectively. 
\label{Fig_poltime}
}
\end{figure}

{
\subsection{Line Polarization in Polar Coordinates}~\label{sec_line_pol}
}

To further visualise the geometric distribution of the \ion{Si}{\sc II} and \ion{Ca}{\sc II} opacities in the ejecta of SN\,2019np, we cast the line polarization into the format of polar plots where the radial axis indicates the velocity across the spectral profile and the angle from the reference direction represents the polarization position angles on the plane of the sky at the corresponding wavelength (introduced by \citet{Maund_etal_2009}, and
see, e.g., \citealp{Reilly_etal_2016, Hoeflich_2017, Stevance_etal_2019}).  Figure~\ref{Fig_polar} presents the polar plots for the \ion{Si}{{\sc II}} $\lambda$6355 and \ion{Ca}{{\sc II}}\,NIR3 lines from days $-$14.5 to $+$14.5. 

On day $-$14.5, relatively highly polarized \ion{Si}{{\sc II}} is present mostly above the photospheric velocity.  The orientation of the Si-rich material appears to be different from the direction of the dominant axis as determined in Section~\ref{sec_quplane} and indicated by the grey sector in the left panel of Figure~\ref{Fig_polar}.  Note that the angular size of the fan-shaped sector represents the 1$\sigma$ uncertainty of the position angle.  Unlike the Si-rich material that is confined in a relatively narrow range in position angle, the Ca-rich component exhibits a more diverse radial profile.  The Ca-rich material below the high-velocity (hv) component at $\sim 20,000$\,km\,s$^{-1}$ shows a range in position angle that is consistent with (i) the dominant axes plotted as black dashed lines in the left panels of Figure~\ref{Fig_qu}, and (ii) the grey fan-shaped sector in the left panel of Figure~\ref{Fig_polar}.  However, the component above the high-velocity threshold exhibits a range in position angle that is distinct from the dominant axis but has a similar orientation as the Si-rich material above the photosphere.  Therefore, the high-velocity Si-rich and Ca-rich components seen on day $-$14.5 are likely to share a similar geometric distribution that differs from that of the optical continuum. 

On day $-$11.4, the dominant axis has rotated relative to day $-$14.5, as indicated by the position angle of the grey fan-shaped sector in the second polar plot of Figure~\ref{Fig_polar}.  Additionally, based on its reduced angular extent, we deduce that the symmetry axis of the SN ejecta becomes more prominent and well-defined as the photosphere progressively recedes.  Most of the Si- and the Ca-rich material gets almost aligned with the optical dominant axis, with larger offsets seen in the radial profile of the Ca-rich component.  This alignment suggests that a similar axial symmetry is shared by the total ejecta and the line opacities.  An overall similar geometry of SN\,2019np can be derived from the polar plots for days $-$6.4 and $+$0.5 (third and fourth panels in Figure~\ref{Fig_polar}), which indicate no significant evolution since day $-$11.4.  From day $-$11.4 to $+$0.5, the orientation of the dominant axis persists.  The widths in velocity of the fan-shaped sectors display an overall decreasing trend for both the Si-rich and the Ca-rich components.  Since the line velocities decrease and the high-velocity components diminish with time, the polarization signal measured at the high-velocity end decreases and becomes less significant as indicated by the large uncertainties. 

By day $+$14.5, the dominant axis has rotated compared to that measured during the rising phase of SN\,2019np.  The scatter has increased \textcolor{black}{again} in  radial profiles of the Ca-rich material, suggesting a more complex structure of the line-forming regions in the more inner layers of the SN ejecta.  The high-velocity component has become indiscernible in the flux spectrum (Figures~\ref{Fig_iqu_ep5} and \ref{Fig_pol}, and \citealp{Sai_etal_2022}).

\textcolor{black}{An overall property of the polar diagrams is their patchy appearance, especially in \ion{Ca}{\sc II}\,NIR3
(Fig. \ref{Fig_pol}). These ``flocculent'' structures tend to become gradually less conspicuous with time, and increase again at day  $+$14.5.}

\section{Numerical Modelling}~\label{sec_model}

This section conducts a quantitative study of the degree of asphericity of SN\,2019np inferred from the observations described in Section~\ref{sec_spec_pol}.  We also investigate their temporal evolution and interpret the nature of the polarization variations on small wavelength scales.  As a baseline, we will use an off-center delayed-detonation model \citep{1991A&A...245..114K}, namely the explosion of an $M_{\rm Ch}$ WD in which a deflagration front starts in the center and transitions to a detonation for reasons described in Section~\ref{sec_ref_model}.

A low level of polarization along the continuum spectrum of a SN is most likely generated by spherically symmetric ejecta leading to complete cancellation of the electric vectors.  However, an aspherical but rotationally symmetric object may also be viewed along its symmetry axis, which has the same effect.  To distinguish these two possibilities, we will use both the polarization over the quasi-continuum and the modulation of the polarization across major spectral features in order to separate the intrinsic asphericity and the polarization actually observed from a certain direction.  In our analysis, we will employ an approach of minimum complexity rather than fine tuning the parameterised geometry to optimise the fitting.  The modeling will address whether the 0.1\%--0.2\% polarization variations with wavelength in the quasi-continuum seen at all epochs can be understood in terms of opacity variations.  Furthermore, we will discuss whether the temporal and spectral resolution of our VLT spectropolarimetry is sufficient to detect and probe any small-scale structures in density and/or abundances. 

The VLT spectropolarimetry of SN\,2019np between days $-$14.5 and $+$14.5 was analysed through simulations employing modules of the HYDrodynamical RAdiation (HYDRA) code\footnote{Many of the HYDRA modules are regularly used by other groups and are available on request to P.H.}
\citep{1995ApJ...440..821H,2003ASPC..288..185H,Penney_etal_2014,2021ApJ...923..210H,2021ApJ...922..186H}. 
HYDRA solves the time-dependent radiation transport equation (RTE) including the rate equations that calculate the nuclear reactions based on a network with 211 isotopes and statistical equations for the atomic level populations, the equation of state, the matter opacities, and the hydrodynamic evolution.  The resulting polarization is obtained by post-processing the given level populations and the density and abundance structure through a Monte Carlo (MC) approach \citep{1991A&A...245..114K, 1995ApJ...440..821H, 2003ASPC..288..185H, Penney_etal_2014, 2021ApJ...923..210H, 2021ApJ...922..186H}.  Atomic models were considered for the ionisation stages I--III of C, N, O, Ne, Mg, Na, Ca, Si, S, Ar, V, Ti, Cr, Fe, Co, and Ni, but without forbidden transitions.  For the luminosity evolution of the multidimensional model as a function of time, a spherical reference model with 911 depth points was adopted, which is adequate considering the small deviation from spherical symmetry.  Moreover, the timescales are dominated by the inner layers which are almost spherical in off-center delayed-detonations whereas the spectra are formed in the photosphere.  This allows us to compare the observations with snapshots of the multidimensional model, neglecting time derivatives in the rate and radiation transport equations. 

For the polarization spectra, we use $\sim 700$ frequency counters between 2800 and 10,200\,\AA. The resulting spatial discretisation corresponds to a formal spectral resolving power of $R \approx 500$, which matches that of the observations ($R \approx 440$, Section~\ref{sec_obs}).  However, in a rapidly expanding atmosphere with gradients, the spatial resolution degrades $R$ to $\sim 150$ since the solution of the radiation transport equation depends on the spatial gradients of the physical quantities.  Simulating a large number of configurations by multidimensional models is prohibitively expensive.  Therefore, we employ a scattering approach with a thermalisation depth to find and discuss estimates for the degree of asphericity in the surface as well as deeper layers \citep{Hoeflich_1991}.

\begin{figure*}
\includegraphics[width=0.9\linewidth]{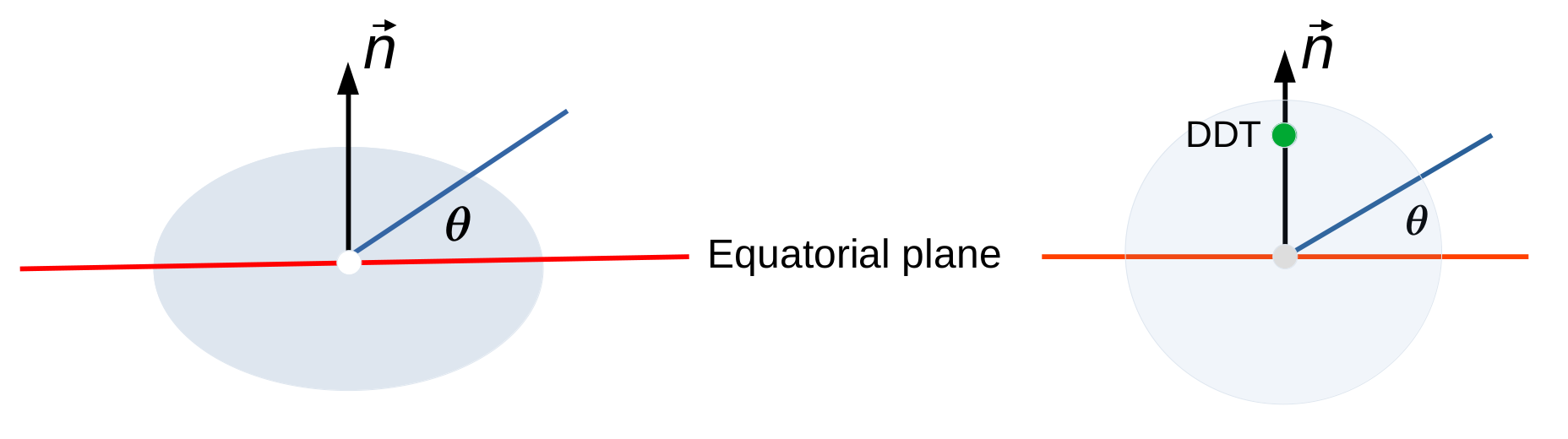} 
\caption{The symmetry axis $\vec{n}$ (black arrow) is defined by the minor axis of a rotationally symmetric ellipsoid (i.e., an oblate spheroid, left plot) or by the vector (right plot) through the center (gray dot) and the location of the DDT (green dot). The equatorial plane E (red) is given by $\vec{n}$\,$\vec{x}$=\,0 
with  $\vec{x}$ spanning E. 
% for all $\vec{x} \in$ E.
%\textcolor{blue}{\bf  DBA:  Why don't you use the normal vector notation with an arrow above the name?}
The viewing angle $\theta$ is the angle between the plane E and the direction to the observer (blue line). $\theta=+90^\circ$, $-90^\circ$, and $0^\circ$ correspond to the north pole, south pole, and the equator, respectively. {As common, $\theta$ is measured  counter-clockwise.} 
%\textcolor{blue} {and the lower-case $\theta$, and I've changed all new texts accordingly.  But, in the figure, this is still pending.  Furthermore, I've replaced "inclination angle " with "viewing angle" to be consistent with the rest of the paper.}
}
\label{illust}
\end{figure*}

\begin{figure*}
\includegraphics[width=0.35\linewidth]{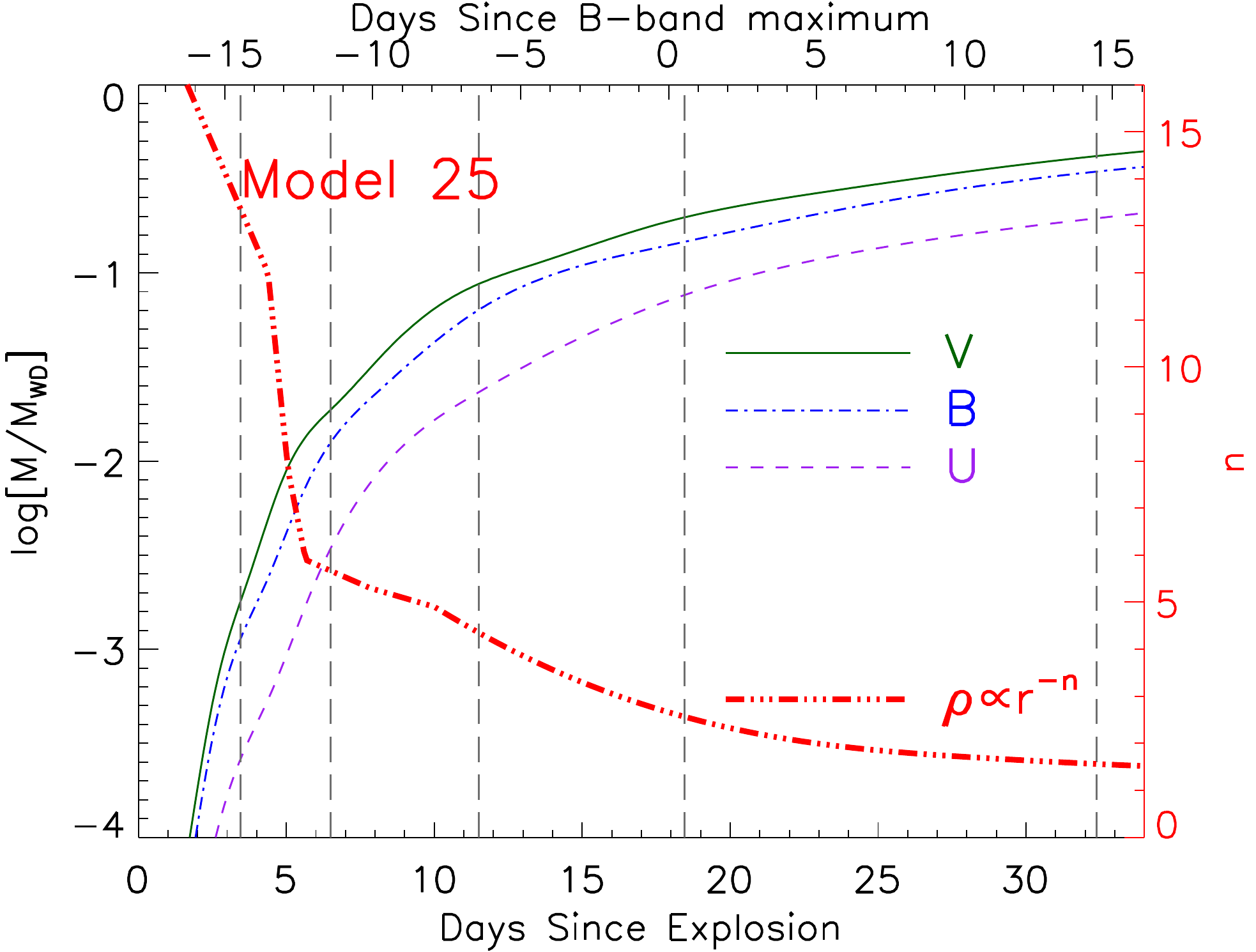}
\includegraphics[width=0.35\linewidth]{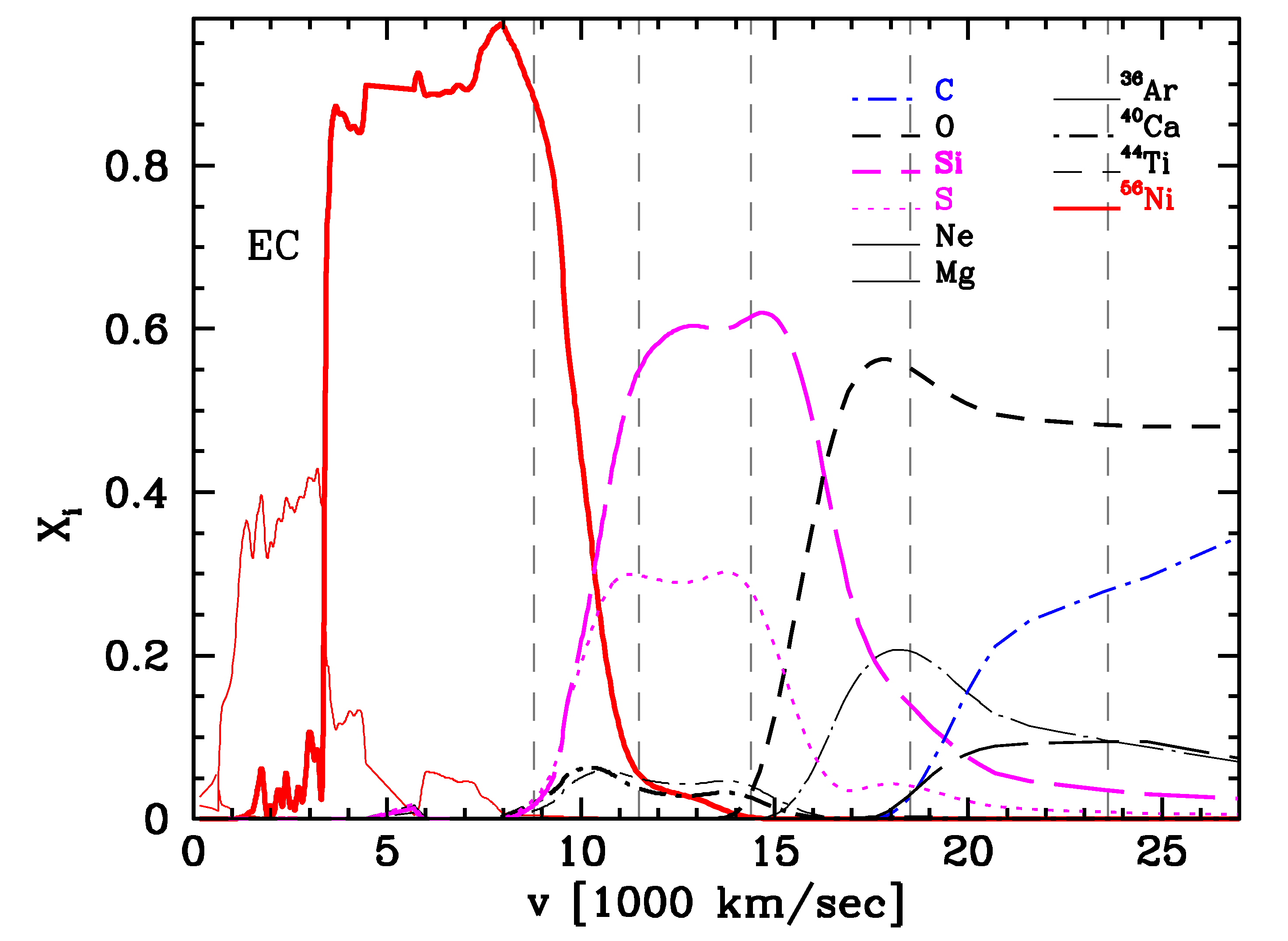}
\includegraphics[width=0.29\linewidth]{off-center07Nin_dot.png}
\caption{{\sl Left:} 
The mass above the photosphere as seen \textcolor{black}{by photons} in the $U$, $B$, and $V$ bands as a function of time for the normal Type Ia SN calculated in the off-center \textcolor{black}{ angle-averaged version of the} delayed-detonation Model 25 \citep{Hoeflich_2017}.  The exponential index ($n$) of the radial density distribution at the position of the photosphere as a function of time is also shown by the red triple-dot-dashed line.  The five epochs of VLT spectropolarimetry are marked by grey vertical dashed lines.  
{\sl Middle:} Angle-averaged abundance structure as a function of expansion velocity, also calculated using Model 25.  Vertical grey-dashed lines indicate the location  of the scattering photosphere --- that is, $\tau_{\rm sc}=1$ at the times when the VLT spectropolarimetry was obtained.  The region with electron-capture elements is indicated by EC. 
{\sl Right:} The $^{56}$Ni distribution as seen above the photosphere on day $+$14.5 based on the hydrodynamical simulation of the off-center detonation. The mass fraction of off-center $^{56}$Ni
\textcolor{black}{above the photospheric radius (dark-red)} 
is $\sim 6$\%.
At this phase, the radius of the photosphere \textcolor{black}{is close to } the location (black dot) where the deflagration-to-detonation transition takes place, \textcolor{black}{ and it expands with $\sim 7000$\,km\,s$^{-1}$}.
The mass fraction is colour-coded in a domain size of $\pm$23,500\,km\,s$^{-1}$.
}
\label{Fig_structure}
\end{figure*}

The continuum polarization may be caused by an aspherical electron-scattering photosphere or an off-center energy input or both \citep{1995ApJ...443...89H, Kasen_2006, Bulla_etal_2016a}
{ (see Fig. \ref{illust})}. 
In the spectra of a Type Ia SN, opacities from bound-bound transitions form a wavelength-dependent quasi-continuum and also produce individual spectral features.  The quasi-continuum may exhibit polarization signals when the sizes of any opacity clumps are comparable to the free mean path of Thomson scattering.  One should keep in mind that, in the high Thomson optical-depth regime ($\tau \gtrsim 3$--4), the continuum polarization in the quasi-continuum will be lower compared to that at $\tau \approx 1$ and reach an asymptotic limit \textcolor{black}{for large optical depths} since any information about asphericity will be blurred by multiple scattering (see, e.g., Figures~1 and 5 of \citealp{Hoeflich_1991}).  If the opacity of the quasi-continuum becomes much larger than the optical depth of the Thomson scattering, the degree of polarization $p \propto \tau_{\rm sc}$, where $\tau_{\rm sc}$ denotes the electron-scattering optical depth of layers at which photons thermalise.

\begin{figure*}
\includegraphics[width=1.0\linewidth]{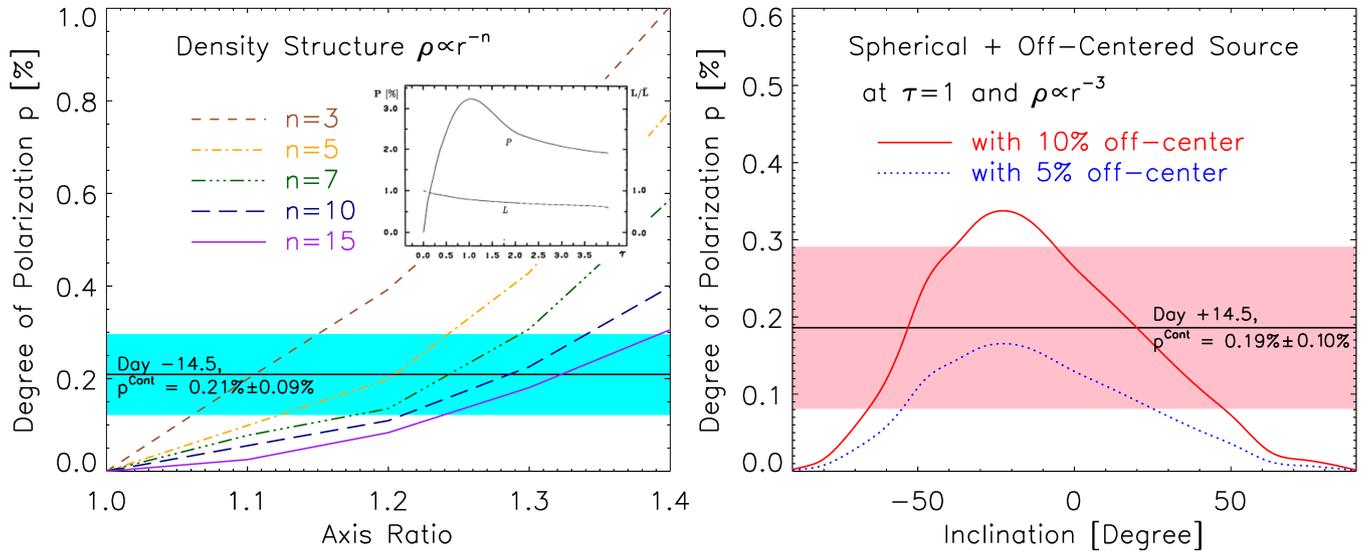} 
\caption{{\sl Left:} Continuum polarization as a function of asphericity for an oblate ellipsoidal scattering-dominated photosphere viewed equator-on.  A steep density gradient can be expected at early phases (see Figure~\ref{Fig_structure}), when the polarization approaches the limit of large optical depth.  The exponential index of the radial density distribution on day $-$14.5 can be best represented by configurations with $n \approx 13$--14 (left plot of Fig. \ref{Fig_structure}).  The horizontal line and the cyan-shaded area indicate the level and the associated uncertainty of the continuum polarization (respectively) measured on day $-$14.5.  Therefore, an axis ratio between $\sim 1.25$ and 1.4 is expected. \textcolor{black}{To aid the discussion in the text,} the inset panel shows the continuum polarization as a function of thermalisation optical depth for an oblate ellipsoid with an axis ratio of 2 and $\rho \propto r^{-2}$ (from \citealp{Hoeflich_1991}).  
{\sl Right:} The degree of continuum polarization produced by a spherical photosphere plus an off-center energy source as a function of viewing angle.  For the illustration of the effect of an off-center energy source, a radial density distribution index $n=3$ and an optical depth $\tau=1$ are chosen to represent the SN photosphere around two weeks after maximum \textcolor{black}{brightness} (see Figure~\ref{Fig_structure}).  The horizontal line and the pink-shaded area mark the level and the error of the continuum polarization, respectively, on day \textcolor{black}{$+$}14.5. 
}
\label{Fig_pol_axis_ratio}
\end{figure*}

\subsection{The Reference Model}~\label{sec_ref_model}

As the \textcolor{black}{spherically symmetric} reference, we adopt the delayed-detonation Model 25 for a normal-bright Type~Ia~SN from \citet{Hoeflich_2017} because it shows light-curve properties very similar to \textcolor{black}{those of} SN\,2019np. The explosion disrupts a WD \textcolor{black}{with mass} close to $M_{\rm Ch}$. Burning starts as a deflagration front near the center and transitions to a detonation by the mixing of unburned fuel and hot ashes \citep{1991A&A...245..114K}. 

The explosion originates from a C/O~WD with a main-sequence mass of 5\,M$_\odot$ \textcolor{black}{as the progenitor star}, solar metallicity, and a central density $\rho_c=2 \times 10^9$\,g\,cm$^{-3}$.  The deflagration--detonation transition was triggered when the density at the front had dropped below  $2.5 \times 10^7$\,g\,cm$^{-3}$ when $\sim0.24$\,M$_{\odot}$ of the material had been burned by the deflagration front.  For the construction of the off-center delayed detonation transition (DDT), we follow the description of \citet{1999ApJ...527L..97L} that has been previously employed \citep{2006NewAR..50..470H,2015ApJ...804..140F,2021ApJ...922..186H}. To terminate the deflagration phase, the delayed-detonation transition is triggered with the mass-coordinate $M_{\rm DDT}$ as an additional free parameter.  The time series of the flux and the polarization spectra were generated without further tuning of the model parameters.  The photometric properties predicted by the spherical model are similar to those measured for SN\,2019np, namely $\Delta m_{15}(V/B) = 1.14/0.68$\,mag (Model 25) and 1.04/0.67\,mag \citep{Sai_etal_2022}.

According to the above prescription, the axial symmetry of the SN \textcolor{black}{model} is defined by the location(s) of the point(s) where the deflagration-to-detonation transition took place. The asphericity in the density distribution near the surface layers was characterised by introducing an additional free parameter when modeling the continuum polarization at the earliest phase.  For the actual implementation see the last paragraph of this subsection.
The symmetry axis that determines the geometric properties of the outermost layers and that \textcolor{black}{which} is defined by the location of the deflagration-to-detonation transition in the inner regions are not correlated with each other, since the latter is stochastic and expected to take place deeper in the WD.  As the DDT is turbulently driven in the regime of distributed burning, its location depends on the ignition process of the thermonuclear runaway, namely multispot or off-center ignition, and \textcolor{black}{initial} magnetic fields.  In contrast to the inner symmetry axis, that of the surface layers is likely determined by the direction of the angular momentum of the progenitor system, i.e., the equatorial plane of a companion or the plane of an accretion disc.
Since the luminosity originates from the energy source that is well below the photosphere in the first few days after the SN explosion, and the outermost layers do not affect the emission at later phases, our simulations treat these two symmetry axes as independent parameters.

In Figure~\ref{Fig_structure}, we present the mass above the photosphere as a function of time (left panel) and the radial distribution of the chemical abundances as a function of expansion velocity (middle panel). {Overall, the exploding envelope has the familiar onion-shell-like structure. However, the onion is no longer spherical but elongated as a result of the off-center DDT \citep[see also Fig.\ 2 in ][]{2021ApJ...922..186H}.}  In the left panel, we also mark the times of the VLT spectropolarimetry with respect to both the estimated time of the explosion and the $B$-band light-curve peak.  The earliest spectropolarimetry to date of any Type Ia SN on day $-$14.5 probes the outermost $\lesssim$ (2.5--3) $\times 10^{-3}$\,M$_{\rm WD}$ layer of the exploding WD, corresponding to a mass of $\lesssim 4 \times 10^{-3}$\,M$_{\odot}$.  As deduced from the red triple-dot-dashed curve, at such an early phase, the exponential index of the radial density distribution is $n \approx 13$--14.  In the middle panel of Figure~\ref{Fig_structure}, we mark the locations of the photosphere at each epoch of the VLT spectropolarimetry in velocity space.  Note that the absorption minimum of, for example, \ion{Si}{{\sc II}} $\lambda$6355 does not measure the expansion velocity at the photosphere but the average projected velocity toward an observer. The difference \textcolor{black}{compared} to the expansion velocity at the photosphere is particularly large in zones with steep density profiles.  Since the photosphere recedes over time, multi-epoch spectropolarimetry can tomographically map out the degree of asphericity at different chemical layers.  As indicated by the middle panel of Figure~\ref{Fig_structure}, the cadence of the VLT observations \textcolor{black}{of SN\,2019np} only provides a resolution in expansion velocity of $\sim 6000$\,km\,s$^{-1}$, at which a discrimination of any structures smaller than several thousand km\,s$^{-1}$ in the radial direction is not possible. For a detailed discussion, see Section~\ref{sec_opportunities}. 

We employ a delayed-detonation model considering the fact that \ion{C}{\sc II} was seen in the first epoch on day $-$14.5 (Figure~\ref{Fig_iqu_ep1}), corresponding to the very outer layers of $\lesssim 4 \times 10^{-3}$\,M$_{\odot}$, making a sub-$M_{\rm Ch}$ explosion an unlikely candidate even for the case of C/He mixtures \citep{Shen_Moore_2014}.  Note that $M_{\rm Ch}$ explosions may have a thin H/He-rich surface layer as a result of the accretion phase but at a significantly smaller mass, (1--5) $\times 10^{-4}$\,M$_{\odot}$ \citep{Hoeflich_etal_2019}, an amount below our numerical resolution. Therefore, we neglect the H/He layer in our simulation.

In Model 25, the early-time spectra originate from the region with incomplete explosive carbon burning and an inward-increasing contribution by explosive oxygen burning \textcolor{black}{(Fig. \ref{Fig_structure})}.  By the time of maximum light, the photosphere is formed in layers of complete oxygen burning and partial silicon burning as indicated by the presence of Ar and Ca.  The emergence of Ar lines in the mid-infrared was predicted by our models.  In SN\,2014J, they were detected by \citet{Telesco_etal_2015}.  At $\sim 2$ weeks after peak luminosity, the spectrum on day $+$14.5 obtained by our last epoch of VLT observations is formed at the interface between partial, distributed silicon burning and with burning to nuclear statistical equilibrium (NSE).  The position of this layer coincides with the location where the DDT has been triggered.  Note that in our simulation the point of the DDT does not lead to a strong refraction wave \citep{Gamezo_etal_2005} as in all spherical delayed-detonation models \citep{1991A&A...245..114K}.  The innermost layers undergo weak reactions under NSE conditions, resulting in the production of electron-capture (EC) elements.  

An asphericity in the outermost layers as indicated by the positive detection of the continuum polarization at the first epoch (see Section~\ref{sec_contpol_obs}) is not produced by our hydrodynamical reference model.  To estimate the degree of asphericity \textcolor{black}{at that early epoch}, we describe the density structure of SN\,2019np \textcolor{black}{by stretching} along the radial direction using an oblate ellipsoid with the axis ratio as a free parameter.  The density and abundance structure are directly taken from our reference model, transforming the distance from the center of an element as $r(m) \Rightarrow r(m,\theta)$ \citep{Hoeflich_1991}.  In the toy models for the continuum developed in Section~\ref{sec_contpol}, we treat the orientation as a free parameter.  For reasons of computational feasibility of the full model, we assume that the symmetry axes of the density and abundances are aligned. 
When the deflagration front \textcolor{black}{has burned} $\sim 0.25$\,M$_\odot$, we trigger the detonation by mixing burned and unburned fuel at $M_{\rm DDT}=0.5$\,M$_\odot$\footnote{ Using the amplitude of the \ion{Si}{{\sc II}} $\lambda$6355 polarization as the criterion, we chose this mass fraction from a set of intermediate models for levels of 0.1, 0.3, 0.5, and 0.9\,M$_\odot$. \textcolor{black}{For the Monte Carlo post-processing to obtain $p$ presented in this paper, a number of particles per resolution element has been used to obtain a statistical absolute error of $\lesssim 0.015\%$.}
 }, 
the so-called Zel'dovich reactivity gradient mechanism \citep{Zel'Dovich_etal_1970}.

\subsection{Continuum Polarization}~\label{sec_contpol}

On days $-$14.5 (the first epoch) and $+$14.5 (the fifth and last epoch), the level of the continuum polarization has been measured as 0.21\%$\pm$0.09\% and 0.19\%$\pm$0.10\%, respectively, both at a $\sim 2\sigma$ level \textcolor{black}{(see footnote in Section \ref{sec_contpol_obs})}.  The former corresponds to the very outer layers and the latter probes the inner layers near the position where the deflagration-to-detonation transition takes place.  Between day $-$11.4 (second epoch) and day $+$0.5 (fourth epoch), {the continuum polarization was consistent with zero within the uncertainties} (see Section~\ref{sec_contpol_obs}).

At early times, the thermalisation depth of the photons emitted by a SN is large (i.e., $\tau \gtrsim 3$), and the polarization degree reaches its asymptotic value (see Figs.~1 and 11 of \citet{1993A&A...268..570H} and \citet{1995ApJ...443...89H}, respectively, and the inset in Figure~\ref{Fig_pol_axis_ratio}).  The maximum polarization degree \textcolor{black}{is} expected when $\tau \approx 1$ \citep{Hoeflich_1991}.  Linear polarization produced by aspherical density structures follows the relation $p \propto sin^2 \theta$, where $\theta$ is the angle between the polar direction and the observer.  As the radial density exponent $n$ is high at the first epoch ($n \approx 13$--14; see left panel of Figure~\ref{Fig_structure}), a {minimum} axis ratio of 1.25--1.4 
%at the viewing angle of SN\,2019np 
can be inferred from the continuum polarization of 0.21\%$\pm$0.09\% on day $-$14.5 (see left panel of Figure~\ref{Fig_pol_axis_ratio}).  
For an equator-on perspective {($\theta = 0^{\rm o}$)}, the high axis ratio implies asphericity in excess of 30\% in the $4 \times 10^{-3}$\,M$_{\odot}$ of the carbon-rich layers in the outermost part of the exploding WD (see Figure~\ref{Fig_structure}).  

Only three days later, on day $-$11.4, the continuum polarization had dropped rapidly to a level consistent with zero.  By contrast, for a constant global asphericity, the degree of polarization would increase with time because (i) the density slope becomes flatter (see the left panel of Figure~\ref{Fig_structure}), and (ii) the thermalisation optical depth decreases to $\sim 1$ as the SN reaches maximum light, when the quasi-continuum opacity in the iron-rich region becomes comparable to, {or larger than,}
the Thomson opacity (see, e.g., Figure~2 in \citealp{Hoeflich_etal_1993}).
%\citealp{Hoeflich_etal_1993,1995ApJ...443...89H}).  
  
Therefore, the rapid decrease in continuum polarization observed in SN\,2019np suggests that the large-scale asphericity in the density structures seen at the earliest phase is limited to the very outer layers.  We find that an additional \textcolor{black}{structural} component is only required at the first epoch.  For all deeper layers, we do not have to impose any asphericity on the density distribution. 
The difference in polarization position angle between the surface and the deeper layers \textcolor{black} {may be attributed to the additional structural component.}

On day $+$14.5, the continuum polarization of SN\,2019np exhibited an increase to 0.19\%$\pm$0.10\%, although the scattering optical depth had decreased significantly well below $\tau_{\rm sc}=1$, where $p \propto \tau_{\rm sc}$, \textcolor{black}{and, hence, a decrease in $p$ may be expected.} As will be discussed in Section~\ref{sec_pol_spec}, the continuum polarization can be understood as a consequence of our off-center DDT Model 25, which produces an aspherical distribution of $^{56}$Ni, and thus an inhomogeneous central energy source. 

{ An off-center energy source is needed because the quasi-continuum dominates the Thomson scattering causing thermalisation at low $\tau_{\rm sc}$ and, thus, only photons with grazing incidence on the outer photosphere get polarized.  The reason is that photons scattered into the direction of their travel, the Poynting vector, are unpolarized whereas light scattered orthogonally to the Poynting vector is 100\% polarized.  Radially traveling photons are most likely to escape when they are scattered along the Poynting vector, whereas grazing-incidence photons can escape most easily when they are radially scattered by $90^{\circ}$.}

A similar \textcolor{black}{increase} of the continuum polarization after maximum light was reported for SN\,2019ein, namely 0.28\%$\pm$0.10\% on day $+$10 and 1.31\%$\pm$0.32\% on day $+$21 \citep{2021MNRAS.503..312M,Patra_etal_2022}.  This rise of the broad-band polarization at late phases was also attributed to the emergence of an aspherical central energy input as the photosphere reaches the Si/Fe interface \citep{Patra_etal_2022}.

As a first step, we quantify the level of the asphericity in the $^{56}$Ni distribution required for a toy model that does not depend on details of the explosion process.  We estimate the amount of off-center $^{56}$Ni at the photosphere relative to the main, symmetric component of the $^{56}$Ni distribution based on previous simulations \citep{Hoeflich_1995}. Motivated by the low continuum polarization between days $-$11.4 and $+$0.5 (see Section~\ref{sec_contpol_obs}), we assume\footnote{As shown in Section~\ref{sec_pol_spec}, we cannot allow  a large-scale density asymmetry in Model 25.} that the low continuum polarization by any global asphericity in the density at the photosphere can be neglected and, based on Model 25, that the off-center source is at about the photosphere (see right panel of Figure~\ref{Fig_structure}). \textcolor{black}{For the toy model, a point-like off-center source at $\tau_{sc} = 1$ in a spherical envelope is assumed to obtain a first-order estimate.}

The relative contribution by the off-center component \textcolor{black}{at day $+$14.5} to the total energy input
at the photospheric level is found to be between 5\% and 10\% (see the right panel of Figure~\ref{Fig_pol_axis_ratio}). Using as reference the axis defined by the center and the location the DDT, a tangential energy source causes a flip in the polarization angle \textcolor{black}{ or, in the $Q$--$U$ diagram, the polarization axis should rotate by $90^{\circ}$\citep{1995ApJ...443...89H}. However,
 only $\sim 70^{\circ}$ are observed relative to the layers seen at day $-14.5$} . Thus, our toy model \textcolor{black}{predicts a difference between the symmetry axes of the outer structural component and the inner layers which causes  a change  in PA in the $Q$--$U$ diagram of } about 20$^{\circ}$ \textcolor{black}{compared to day +0.5 (Fig. \ref{Fig_qu})}.  This estimate is obtained by the vectorial addition of the polarization contributions by the off-center source and the spherical source.

Although a change in position angle from day $-$14.5 to $-$11.5 to $+$14.5 is hard to measure in SN\,2019np owing to the very low intrinsic continuum polarization, we suggest that the rotation of the dominant axis fitted to the same optical wavelength range as above (see Figure~\ref{Fig_qu}) is compatible with the prediction of an off-center energy source beginning to be exposed to the observer at this phase. 

One of the major effects of the very steep density slope in the outer layers is that even the small $\sim 0.2$\% continuum polarization $\sim 3.5$ days after the SN explosion implies a significant aspherical density distribution.  The polarization of SN\,2019np at the earliest epoch is higher than that measured in other Type Ia SNe at later phases, which are closer to epoch 2 on day $-$11.4 and epoch 3 on day $-$6.4 of SN\,2019np.  For example, 0.10\%$\pm$0.07\% was observed in SN\,2019ein \citep{Patra_etal_2022} on day $-$10.9, and 0.06\%$\pm$0.12\% in SN\,2012fr \citep{Maund_etal_2013} on day $-$11.  

However, the 0.21\%$\pm$0.09\% continuum polarization of SN\,2019np on day $-$14.5 ($\sim 3.5$ days after the explosion) is comparable to the marginal detection of a 0.20\%$\pm$0.13\% continuum polarization in SN\,2018gv on day $-$13.5, $\sim 5$ days after the explosion, \citep{Yang_etal_2020}.  At that moment, the density exponent in SN\,2018gv had dropped to $\sim 9$--10 (see the left panel of Figure~\ref{Fig_pol_axis_ratio} and Figure~21 of \citealp{Yang_etal_2020}).  This leads to a $\sim 10$\%–-35\% deviation from spherical symmetry within the outermost $\sim$ (0.5--2) $\times 10^{-2}$\,M$_{\rm WD}$ for an equator-on configuration.  The cases of SNe\,2019np and 2018gv may provide a hint that any asphericity in the outer layers of normal-bright Type Ia SNe becomes apparent in polarization only during the very earliest phases and thereafter quickly almost vanishes.  Therefore, given the high density gradient near the surface layers of the ejecta of Type Ia SNe, a low but nonzero continuum polarization measured in the first few days after the explosion does not necessarily imply a low deviation from sphericity in their outermost layers. 

\begin{figure}
\includegraphics[width=1.0\linewidth]{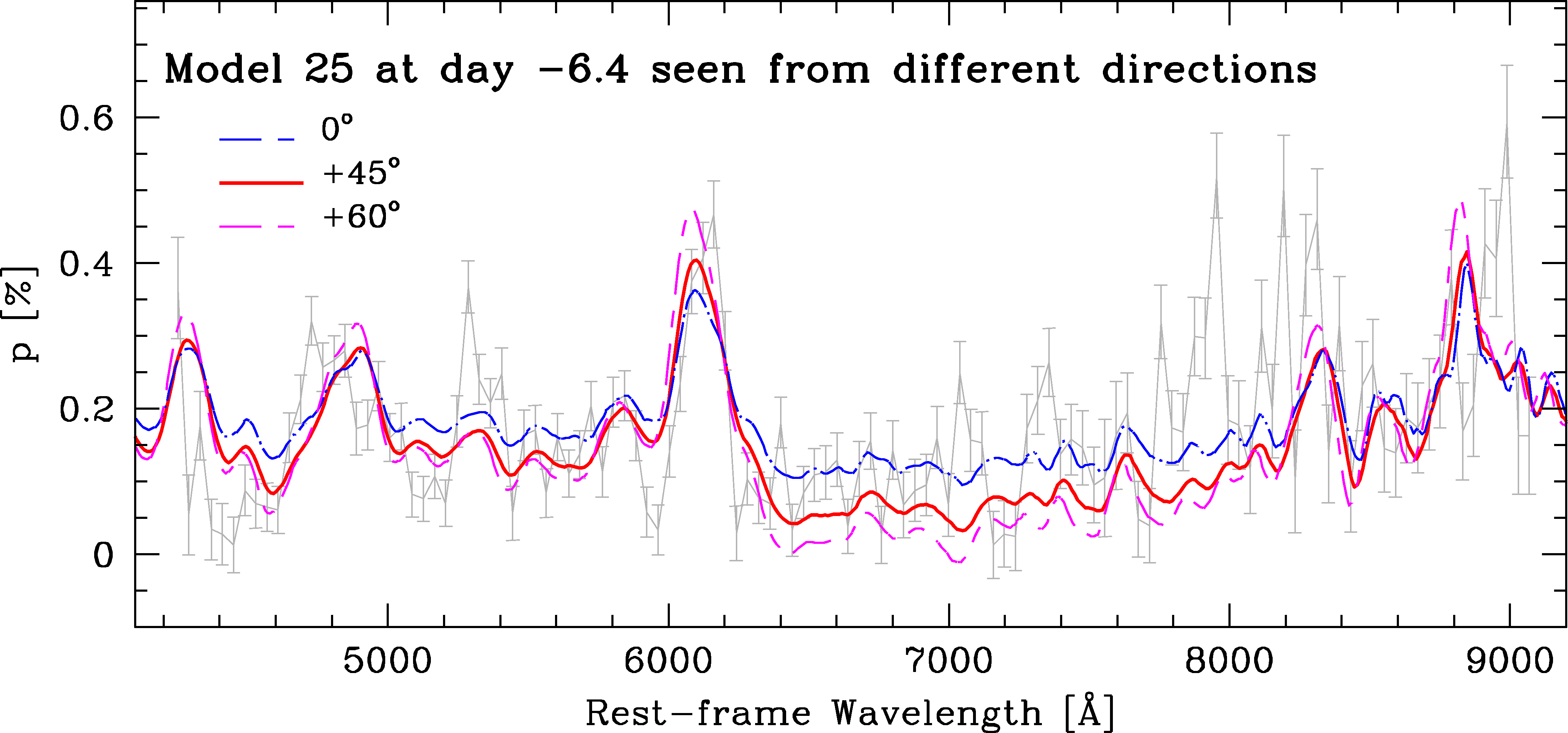}
\caption{{Estimating the viewing angle $\theta$ from polarization spectra (see text). As examples, the full polarization spectrum of the off-center DDT Model 25 at day $-6.4$ is shown for inclinations $\theta$ of about $0^\circ$ (blue, equator-on), $45^\circ$ (red), and $60^\circ$ (cyan), and, as reference, the intrinsic polarization of SN\,2019np at a corresponding resolution (gray).  The estimate of $\theta$ is based on all epochs, and the error in $\theta$ is constrained by the uncertainty of the observations.
{For $\pm 90^{\rm o}$ the polarization is close to zero.}
Note the sensitivity of  the ratio between line and continuum polarization to $\theta$. 
}  
\label{Fig_angle}
}
\end{figure}

\begin{figure*}
\includegraphics[width=1.0\linewidth]{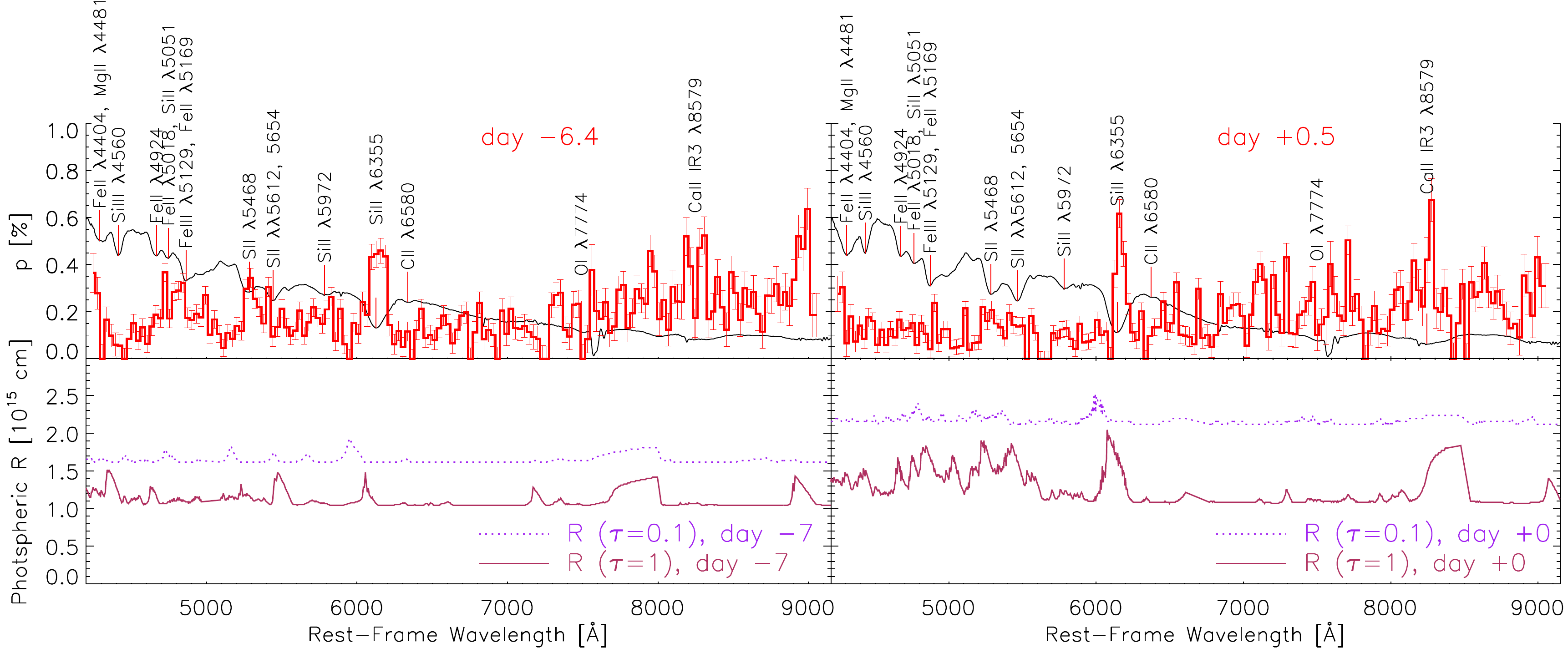}
\caption{Flux and polarization spectra observed in SN\,2019np compared to the spectral formation radius computed with Model 25.
{\it Top panels:}  Scaled flux spectrum (black curve) and degree of polarization (red histogram) observed on days $-$6.4 (left), and $+$0.5 (right). 
{\it Bottom panels:}  Photon-decoupling radius $R$ for $\tau = 0.1$ and 1 computed with delayed-detonation Model 25 at similar phases (left panel for day $-$7 and right panel for day $+$0). Major spectral lines are labeled.  Note that the polarization is mostly produced by Thomson scattering between $\tau= 0.1$ and 1.  Even strong features such as \ion{Si}{{\sc II}} $\lambda$6355, the \ion{Ca}{{\sc II}}\,NIR3 complex, and various blended features below 5500\,\AA\ are formed in the same region, setting a qualitative limit to the picture of line polarization being produced by chemically selective blocking of an underlying scattering-dominated photosphere. 
\label{Fig_tau_max}
}
\end{figure*}

\begin{figure}
\includegraphics[width=1.0\linewidth]{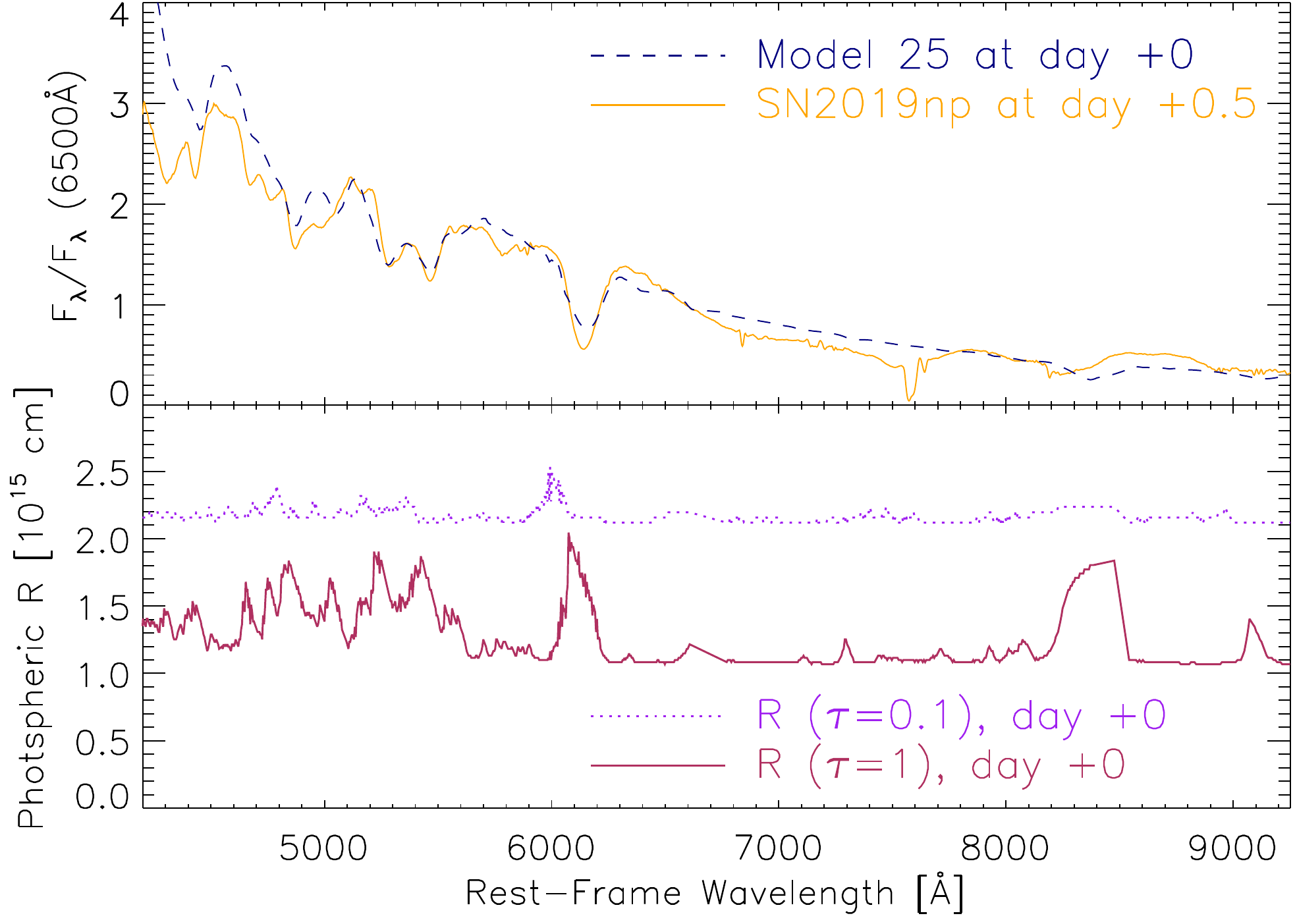}
\caption{Spectral formation as a function of radius at maximum light.  We compare the observation with the spectrum of Model 25 (upper panel) and the radii corresponding to optical depths of 0.1 and 1, between which the absorption features and the polarization are formed (Figure \ref{Fig_pol_axis_ratio}).  The observed and synthetic spectra generally agree without fine tuning.  For strong lines like \ion{Si}{\sc II} $\lambda$6355, the flux minima and the blue wings typically correspond to an optical depth of 1 and 0.1, respectively.  This is also true for other strong features such as \ion{Ca}{\sc II}\,NIR3 at $\sim 8200$\,\AA\ and many of the line blends below 5500\,\AA.  Strong spectral features form in the same region where the continuum polarization is also produced.  The high-frequency structure in the radius of formation of \ion{Si}{\sc II} $\lambda$6355 is due to \ion{Fe}{\sc II} transitions which do not appear in the flux spectra, but show up in the line profiles as discussed in Figure~\ref{Fig_pol_model}.  Note the reduced effect of the quasi-continuum, justifying the wavelength range used for the determination of $p^{\rm cont}$ in Section~\ref{sec_contpol_obs}.
\label{Fig_tau_max2}
}
\end{figure}

\begin{figure}
\includegraphics[width=1.0\linewidth]{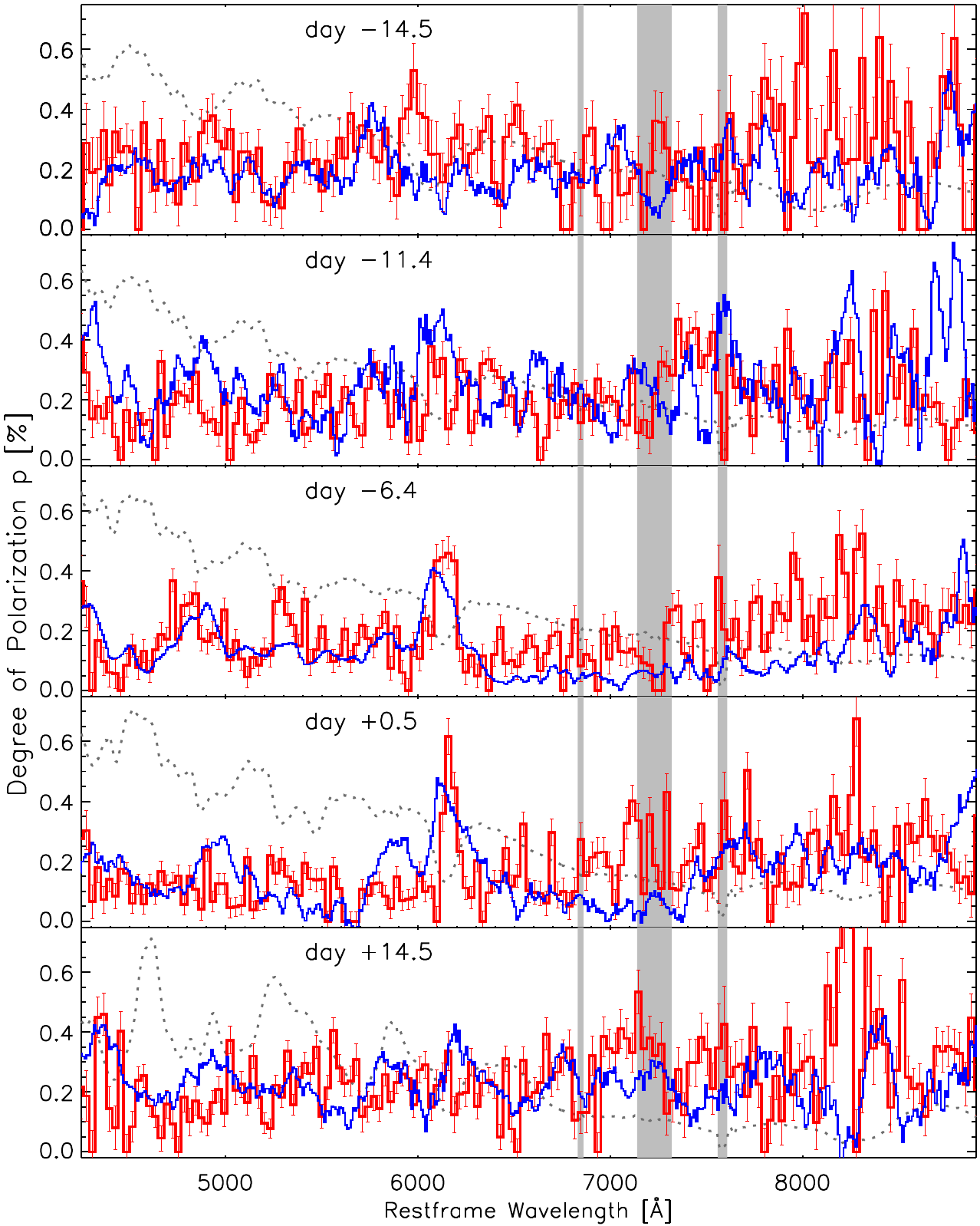}
\caption{Polarization calculated with off-center DDT Model 25 (blue histograms) compared to the intrinsic polarization of SN\,2019np (red histograms) from days $-$14.5 to $+$14.5.  The grey dotted curve in each panel shows the observed scaled flux spectrum at the given epoch.  In the model calculations, an inclination of $\sim 45^{\circ}$ was adopted. 
~\label{Fig_pol_model}
}
\end{figure}

\subsection{Polarization Spectra}~\label{sec_pol_spec}
\label{Sec_pol_spec}

The spectral evolution of SN\,2019np is similar to that of other normal-bright Type Ia SNe \citep{Sai_etal_2022}, enabling us to compare the polarization spectra of SN\,2019np with the models for normal Type Ia SNe discussed by \citet{1995ApJ...440..821H}.  The inclination was obtained by comparing the direction-dependent synthetic polarization spectra to those of SN\,2019np at all epochs and minimising the $\chi^2$ averaged over 100\,\AA-wide bins.  To fit the evolution of the polarization spectra (see Section~\ref{sec_line_pol} and Figs.~\ref{Fig_pol} and \ref{Fig_poltime}), we find that a viewing angle of $\theta = +$45$^{\circ} \pm 10^{\circ}$ is the most plausible approximation of the actual case\footnote{\textcolor{black}{For example, at maximum light, compared to $\theta =+$45$^\circ$, the polarization in \ion{Si}{{\sc II}} $\lambda$6355 is larger by $\sim +$50$\% $ at $\theta$ $\approx +$35$^\circ$, vanishes at $+$90$^\circ$, and becomes small for negative angles depending on the phase, whereas the overall level of $p$ peaks at $\theta \approx +$10$^\circ$.}}.  In Figure~\ref{Fig_pol_model}, the off-center delayed-detonation model viewed at this angle is in good overall agreement with the observations of SN\,2019np.

In Model 25, the polarization across \ion{Si}{{\sc II}} $\lambda$6355 is formed within an extended geometrical structure between 9000 and 27,000\,km\,s$^{-1}$   which undergoes complete and incomplete oxygen burning in velocity (middle panel of Fig.~\ref{Fig_structure}, where the velocity of the photosphere at the time of the observations is indicated).  In the outermost region of partial explosive oxygen and carbon burning, the polarization of \ion{Si}{{\sc II}} $\lambda$6355 is weaker since its abundance diminishes with increasing velocity.  At early times, this line forms close to the region with $\tau_{\rm sc}=1$.  Because the polarization by electron scattering is mostly formed in the range $0.1 \textless \tau_{\rm sc} \textless 1$, the polarization across \ion{Si}{\sc II} lines is generally low.  The Si polarization increases as the photosphere continuously recedes \textcolor{black}{and, without the structural component, reaches its peak} when the photosphere enters the layers with quasi-equilibrium conditions around the Si group.  Thus, the polarization in \ion{Si}{{\sc II}} $\lambda$6355 increases with growing distance between the optical depth at a given wavelength and the layer with $\tau_{\rm sc}\approx 1$, which is always more internal.  Any aspherical distribution is expected to be most prominent around this phase, when the photosphere passes the inner boundary of the explosive C- and O-burning, and the QSE(Si)/NSE interface becomes exposed \citep{Hoeflich_etal_2006}. 
After peak luminosity, the polarization of \ion{Si}{\sc II} $\lambda$6355 decreases because the quasi-continuum opacities increasingly dominate the electron scattering.

Apart from \ion{Si}{{\sc II}} $\lambda$6355, the polarization over a quasi-continuum wavelength range also increases in the same region that forms various other spectral features, which are resolved
\textcolor{black}{
(see Figures~\ref{Fig_tau_max}, \ref{Fig_tau_max2}, and \ref{Fig_pol_model})}.  This concerns the entire wavelength range occupied by blends of the Fe group, \ion{Si}{{\sc II}}, \ion{S}{{\sc II}}, and \ion{O}{\sc I}.  Depending on time, features \textcolor{black}{in the polarization spectra} appear around (for example) 4400, 4800, 5400, 5800, 6800, 7200, 7500, 8300, and 9000\,\AA\ (Figure~\ref{Fig_pol_model}).  Overall, thousands of overlapping lines are involved (see, e.g., Figures~1 and 2 of \citealp{1993A&A...268..570H}).  The variations in this quasi-continuum depend on the velocity gradients, the abundances, and the ionisation level.  The polarization is very sensitive to this pattern as it influences the thermalisation optical depth by individual components because spectral lines mostly depolarize.  In flux spectra, these variations are mostly blurred because photons are absorbed and emitted, but they are visible in the line-formation radii 
traced by spectropolarimetry (Fig.\ref{Fig_tau_max2}).  As a result, spectropolarimetry is effectively much more sensitive to spectral lines than flux spectroscopy because its observable signatures are much less volatile. 
Nevertheless, some individual patterns can be identified by comparing the line identifications (Figures~\ref{Fig_iqu_ep1} to \ref{Fig_iqu_ep5} and \ref{Fig_tau_max2}) and, from the models, by variations in the wavelength-dependent radii of line formation as presented in Figure~\ref{Fig_tau_max}. For instance, the rather persistent feature at 9000\,\AA\ can be attributed to a strong \ion{Fe}{{\sc II}} + \ion{Co}{{\sc II}} blend which becomes obvious as a change in the thermalisation radius and appears in both observed and synthetic spectra (Figure~\ref{Fig_pol_model}).

In the models, a maximum or minimum in polarization is produced if the thermalisation optical depth is above or below $\tau_{\rm sc} \approx 1$, respectively. \textcolor{black}{Owing} to the sensitivity of the polarization, maxima in the observations can be minima in the synthetic spectra for moderately strong blends which appear in the optical depth (Figure~\ref{Fig_tau_max}). One example is the Fe/Co blend at $\sim 9000$\,\AA. This feature toggles with time between maxima and minima in both theory and observations.  At most epochs, the changes in model and observations are synchronised, except for days $-$11.4 and $+$0.5.  Another example is the S/Fe blend at $\sim 5800$\,\AA, for which the simulations mostly reproduce the observations but on day $+$0.5. Both features can be identified as an elevation in the radius of photon decoupling as shown by Figure~\ref{Fig_tau_max}. For these examples, the time to toggle from high to low polarization can be estimated from the rate with which the photosphere recedes over the abundance gradients.  The gradients typically extend over $\sim 500$--1000\,km\,s$^{-1}$ (Figure~\ref{Fig_structure}) and so correspond to a timescale of $\sim 1$ day. From spectral analysis, a similar timescale of a few days is well established for changes in ionisation stages \citep{1981ApJ...244..780B,1993A&A...269..423M,1995ApJ...443...89H,2001ApJ...557..266L,2006ApJ...645..480B,2014MNRAS.441.3249D}. This is faster than our observing frequency of SN\,2019np and may reflect an insignificant phase shift in the evolution of the models relative to the observations. This phase shift may also point toward small-scale structures such as Rayleigh-Taylor fingers or Kelvin-Helmholtz instabilities not included in our models, which would reveal themselves in short-term variations in the polarization spectra, but are not resolved in our dataset. { The numerous wiggles (Section~\ref{sec_quplane}  and Figure~\ref{Fig_pol_model}) are not resolved in the current observations.  They may possibly be understood in the same way as resolved \textcolor{black}{features:} \textcolor{black}{namely in terms of atomic physics}.  Many coincide with features produced by the model.  Some of them may indicate genuine small-scale structures, and others may be just noise in the data or ejecta.  Their true nature cannot be determined with the current observations.}
This ambiguity points toward a need for high-cadence observations \textcolor{black}{to separate small scale instabilities from imprints governed by atomic physics.}

The change of the polarization profiles of \textcolor{black}{(for example)} \ion{Si}{\sc II} $\lambda$6355 can also be understood within the same framework, leading to a new diagnostic (Figures~\ref{Fig_tau_max} and \ref{Fig_pol_model}) of substructures in lines, although the spectral resolution of our polarimetry may not be sufficient to fully reveal the underlying velocity structure.  Overall, the location of the peak (its Doppler shift) agrees between observations and synthetic profiles including the evolution of the line width. This supports the interpretation by large-scale asphericity in the abundance distributions.  From the models, this evolution can be understood even though, at higher granularity, some discrepancies need to be discussed.  (a) Binning of the data may introduce artifacts, in particular at very early times when the \textcolor{black}{SNR} is low in the current data as on day $-$14.5\footnote{\textcolor{black}{On day $-$14.5, \ion{Si}{\sc II} $\lambda$6355 shows multiple components at 25\,\AA\ binning.}} or on day $+$0.5, when the polarization peak in 
{ \ion{Si}{\sc II} $\lambda$6355}
occupies just one wavelength bin whereas the associated change in the position angle takes place over three bins (Figure~\ref{Fig_iqu_ep4}). 
 (b) Some discrepancies between observations and model profiles may also hint at the model Fe opacities being too weak between 6000 and 6500\,\AA, possibly owing to a lack of Rayleigh-Taylor mixing, too low excitation of the atomic levels, or slightly too low a metallicity in the progenitor.  As discussed above, even the strong lines are blended with many weak lines, which do not appear in the flux but in the polarization. 

In the simulations, the strength of the polarization depends on the thermalisation depth in the atmosphere and the density profile (see Section~\ref{sec_contpol}). If at some wavelength the thermalisation depth is close to the Thomson optical depth of 1, the polarization peaks at that wavelength.  It becomes smaller with decreasing thermalisation depth, and reaches the asymptotic value for large depths.  As a result, the line profile is broader at early times when, owing to steep density profiles combined with decreasing abundances in the region of incomplete oxygen burning, namely around day $-$11.4, the radii of the photon decoupling regions are similar and, thus, the resulting profiles are broad.  With time, the density slope flattens and, to first order, the profile becomes narrower.  Note that, in an expanding atmosphere, the absorption is determined by the Sobolev optical depth which is not inherently spherically symmetric \textcolor{black}{in wavelength} (see Figure~\ref{Fig_tau_max}). By days $-$6.4 and $+$0.5, the Si profile is formed in the QSE region and a flat density gradient leads to an increasing blueshift of the peak.

On day $-$11.4, \ion{Si}{\sc II} $\lambda$6355 is blended with \ion{Fe}{\sc II} $\lambda\lambda$6293, 6358, 6497 and weaker \ion{Fe}{\sc II} and \ion{Fe}{\sc III} transitions from excited levels, leading to a more complicated profile.  In the models, the iron blends seem to be weaker than in the observations. Because line absorption depolarizes, this can explain the lack of depolarization in the model profile in both the blue and red.  For the same reason, at days $-$6.4 and $+$0.5, the observed profile has a steep decline whereas the synthetic profile shows a long red tail.  The imprint of \ion{Fe}{\sc II} $\lambda$6497 may be seen in the observations.  A similar shape of the \ion{Si}{\sc II} $\lambda$6355 polarization profile has also been seen in SN\,2018gv around peak luminosity \citep{Yang_etal_2020}.

In the models and the observations after peak luminosity of SN\,2019np, \ion{Si}{{\sc II}} $\lambda$6355 becomes progressively blended with several strong \ion{Fe}{{\sc II}} lines.  The Si line has vanished by day $+$14.5 as the photosphere recedes into the NSE region, which displays strong Fe-group elements that form both a quasi-continuum and discrete lines in the spectrum.  The feature at $\sim 6200$\,\AA, which is conventionally attributed to \ion{Si}{{\sc II}} $\lambda$6355, becomes increasingly dominated by \ion{Fe}{{\sc II}} lines and, as a consequence, the corresponding polarization across this wavelength range also disappears.  Overall, the quasi-continuum on day $+$14.5 is produced by numerous overlapping Fe-group lines from Fe, Co, Ni, etc. 

Without fine-tuning of our model, the polarization spectra across several major spectral features can also be reproduced and generally agree with the observed polarization spectra (see Figure~\ref{Fig_pol_model}).  For the calculation of the continuum polarization, we use the wavelength range 6400--7000\,\AA\ applied in Section~\ref{sec_contpol_obs}. The global asphericity in the electron density distribution on day $-$14.5 is not accounted for in the hydro simulation.  Therefore, we imposed an overall elliptical distribution with an axis ratio of 2 to obtain the overall level of the polarization over the entire wavelength range observed (see Section~\ref{sec_ref_model}).  The choice of the axis ratio is motivated by Figure~\ref{Fig_pol_axis_ratio} and results quantitatively from the most likely viewing angle, $\theta \approx 45^{\circ}$ (Section~\ref{sec_pol_spec}) and the relation $p \propto {\rm sin}^2 \theta$ (Section~\ref{sec_contpol}).  

At all later epochs, the continuum polarization is calculated directly without modifying the hydro model (Section~\ref{sec_ref_model}).
The continuum polarization produced by the detailed off-center model (Figure~\ref{Fig_pol_model}) is within the 1$\sigma$ error range of the observed values (Table~\ref{Table_pol}).  On day $-$11.4, the asphericity in the electron density is caused by the aspherical abundance distributions.\footnote{\textcolor{black}{In SNe~Ia and unlike SNe~II, the resulting asphericity in the electron distribution remains rather small, 5--10\%, because the free electrons per nucleon  are about equal for Si/S II and Fe/Co II-III.} \textcolor{black}{For example, in the hydrogen-rich envelope of SNe~IIP, the opacity drops by 4 orders of magnitudes over the recombination front of hydrogen, causing 
highly aspherical Thomson-scattering dominated photospheres even in case of slightly aspherical $^{56}$Ni distributions or rotation \citep{2001AIPC..586..459H,2005ASPC..342..330L}. }} Its value is ill-defined because of steep changes of the synthetic polarization at the edges of the 6400--7000\,\AA\ wavelength range \textcolor{black}{ (see second panel from top of Figure \ref{Fig_pol_model})}, \textcolor{black}{ although we used the same range as in Section \ref{sec_contpol_obs} to minimise the effect of lines.}
Therefore, the value of 0.16\%$\pm$0.04\% returned by the models is somewhat larger than the observed level of 0.099\%$\pm$ 0.080\%, but well within the error range. The synthetic continuum polarization on days $-$6.4, $+$0.5, and $+$14.5 are $0.05\% (0.075\pm 0.080\%$), $0.06\%$ $(0.110\pm 0.100\%)$, and $0.22\%$ ($0.186\pm 0.104\%)$, respectively, with the observed values given in brackets\footnote{\textcolor{black}{Note that the variations in the observed continuum polarization are on a 2$\sigma$ level 
(Sect. \ref{sec_isp}). However, they also coincide with a change in the dominant axis in the $Q$--$U$ diagram (Fig. \ref{Fig_qu}), and $\sigma$ includes spectral variations by lines. }}.

Most of the discrete polarization features are at the level of $\lesssim 0.2$--0.4\%.  In the models, they are produced by depolarization or the frequency variation in the thermalisation optical depth.  Whether they appear as local maxima or minima in the polarization spectrum depends on the scattering optical depth of the corresponding region of formation.  Many of these wiggles in the observed polarization spectra (Section~\ref{sec_quplane}) coincide with features in the synthetic polarization spectra.  Discrepancies may be due to small-scale structures that the observations of SN\,2019np do not resolve in time and wavelength.  This may hint at the possibility of significant detection of numerous weakly polarized lines in future higher \textcolor{black}{SNR} observations with FORS2 at ESO's VLT.  Since some patterns do not have mutual counterparts, such observations should also aim for higher spectral resolution.  
Simulations \textcolor{black}{with matching resolution} are feasible with moderate additional effort (see Section~\ref{sec_opportunities}).  

Although our models reproduce many (even relatively minor) aspects of the observations, some limitations are also apparent.  For instance, on day $-$14.5, the synthetic polarization spectra exhibit fairly similar overall patterns, but the models do not show the large-amplitude fluctuations with wavelength observed in the polarization spectra of the strongest resonance lines, namely in \ion{Ca}{\sc II}\,NIR3 (see Figure~\ref{Fig_polar} and Section~\ref{sec_line_pol}) and possibly in \ion{Si}{{\sc II}} $\lambda$6355.\footnote{On day $-$14.5, the feature at $\sim 6000$\,\AA\ (within the \ion{Si}{\sc II} $\lambda$6355 profile, Figure~\ref{Fig_pol_model}), which occupies a single \textcolor{black}{bin} with 30\,\AA\ binning, breaks up into two components with peaks at 0.48\% and 0.39\% if 40\,\AA\ bins are used.}
The involvement of Ni, Co, or Fe seems to be ruled out because, in the spectral region strongly affected by iron-group elements ($\lambda \lesssim 4700$\,\AA\ and $\lambda \gtrsim 8500$\,\AA), similar patterns do not exist and observations and synthetic spectra agree well for our model with solar abundances in the outer C/O layers and Fe, Co, and Ni mass fractions of 0.002, 0.00005, and 0.0001, respectively \citep{1989GeCoA..53..197A}.  As discussed in Section~\ref{sec_opportunities}, an amount of $M(^{56}{\rm Ni}) \approx 0.02$--0.03\,M$_\odot$ in the outer 0.2\,M$_\odot$ ($\sim 0.1$ in mass fraction; alternatively produced in a sub-$M_{\rm Ch}$ explosion) is needed to explain the early bumps in light curves of SNe\,2017cbv and 2018oh.  The \textcolor{black}{associated} spectra are dominated by Fe, Co, and Ni lines \citep{1998sese.conf..693H,2022MNRAS.642.189M}, traces of which are likely just barely seen in all spectra of SN\,2019np (Figure~\ref{Fig_pol_model}). 

Another example \textcolor{black}{for the shortcomings of our current model} is the increased polarization in \ion{Ca}{\sc II}\,NIR3 around and after maximum light, which is not reproduced by our models.  This resonance line has by far the largest cross-section and is optically thick even in regions with solar abundances, so that very minor inhomogeneities can have a big impact on the polarization.  Extended, \textcolor{black}{inhomogeneous} radial components \textcolor{black}{in the Ca distribution} may be expected from Rayleigh-Taylor instabilities, interactions with a companion star, and/or sheet-like/caustic structures, which may develop within 5--10 days after the explosion as the result of mixing of radioactive $^{56}$Ni and electron-capture elements \citep{2000ApJS..128..615M,Fesen_etal_2007,Hoeflich_2017,2018ApJ...861...78M}.  Additionally, the late rise of the \ion{Ca}{\sc II}\,NIR3 polarization may also be caused by the alignment of calcium atoms in the presence of a magnetic field as recently suggested by \citet{Yang_etal_2022_21rhu}.  If combined with near- and mid-infrared nebular spectra, later-epoch polarimetry of the \ion{Ca}{\sc II}\,NIR3 feature will allow us to discriminate between various possibilities concerning the nature of the progenitor and the explosion mechanism as discussed by \citet{Hoeflich_etal_2004}, \citet{Telesco_etal_2015}, \citet{2021ApJ...922..186H}, and \citet{2021jwst.prop.2114A}, since the spatial distribution of radioactive Co and stable Fe, Ni, and Co can be probed independently.

\subsection{SN\,2019\lowercase{np} in Polarimetric Context with other Type Ia SNe}~\label{sec_other_sne}

The polarization properties of some Type Ia SNe are remarkably different from those that are typical for normally bright thermonuclear SNe \citep{Cikota_etal_2019,Patra_etal_2022} and SN\,2019np. An example is SN\,2004dt, which exhibited exceptionally high polarization in some spectral lines.  For instance, the peak polarization across \ion{Si}{\sc II} $\lambda$6355 reached $\sim 2.4$\% and $\sim 3$\% after binning to 50\,\AA\ and 25\,\AA, respectively \citep{Wang_etal_2006_04dt, Cikota_etal_2019}.  Although the continuum polarization was as low as $\lesssim 0.2$\%--0.3\% around peak brightness \citep{Leonard_etal_2005, Wang_etal_2006_04dt}, many features of Si, S, and Mg in the synthetic polarization spectrum\footnote{The continuum polarization in the off-center DDT is $\sim 0.1$\%--0.2\%.} had their equivalent in the observations \citep{2006NewAR..50..470H} as commonly found in Type Ia SNe.  These findings might be accounted for by either a violent merger of two \textcolor{black}{C-O} WDs \citep{Bulla_etal_2016a} or an off-center delayed detonation model within a continuum of parameters \textcolor{black}{consisting} of especially the position of the delayed-detonation transition, the amount of burning during the deflagration phase, and the viewing angle of the observer \citep{Hoeflich_etal_2006}.  

For normally bright Type Ia SNe, the polarization of \ion{Si}{\sc II} $\lambda$6355 five days before $B$-band maximum light ($p^{\rm -5\,d}_{\ion{Si}{\sc II}}$), which is representative of the maximum value ($p^{\rm max}_{\ion{Si}{\sc II}}$), correlates with the light-curve stretch parameter measured as the decline in magnitude within 15 days after $B$ maximum ($\Delta {\rm m}_{15}(B)$; \citealp{Wang_etal_2007, Cikota_etal_2019}).  
For SN\,2019np, \citet{Sai_etal_2022} measured $\Delta {\rm m}_{15} (B) = 1.04 \pm 0.04$\,mag, and $p^{\rm max}_{\ion{Si}{\sc II}}$ amounted to 0.62$\pm$0.03\% on day $+$0.5 (with 30\,\AA\ binning; Figure~\ref{Fig_poltime}).  Therefore, we conclude that SN\,2019np is consistent with the $p^{\rm max}_{\ion{Si}{\sc II}}$--$\Delta {\rm m}_{15} (B)$ relation. 

In a follow-up study, \citet{Maund_etal_2010} investigated \ion{Si}{\sc ii} $\lambda$6355 observations of a sample of nine normal Type Ia SNe and found that $p^{\rm -5\,d}_{\ion{Si}{\sc II}}$ is also correlated with the temporal velocity gradient $\dot{v}_{\rm \ion{Si}{\sc II}}$.  Accordingly, the deceleration of the SN expansion is also correlated with the degree of chemical asphericity. The interpolated velocity gradient of SN\,2019np was 21$\pm$5\,km\,s$^{-1}$\,day$^{-1}$ on day $+$10 \citep{Sai_etal_2022}.  By interpolating the \ion{Si}{\sc II} $\lambda$6355 velocity evolution estimated from our VLT observations, we estimated velocity gradients of 53$\pm$19 and 22$\pm$9\,km\,s$^{-1}$\,day$^{-1}$ on days $+$0 and $+$10, respectively. This means that SN\,2019np was also consistent with the $p^{\rm -5\,d}_{\ion{Si}{\sc II}}$--$\dot{v}_{\rm \ion{Si}{\sc II}}$ relation. 

Subluminous Type Ia SNe exhibit substantially different polarization properties than discussed above.  For instance, a polarization of $\sim 0.7$\% in the optical continuum but only $\sim 0.3$\% across \ion{Si}{\sc II} $\lambda$6355 were observed in SNe\,1999by \citep{Howell_etal_2001} and 2005ke \citep{Patat_etal_2012} at $\sim 0$ and $-$7 days relative to maximum light, respectively.  The high degree of continuum polarization can be explained by a global asphericity of as much as 15\% \citep{Patat_etal_2012}.  These two events are outliers from the correlation proposed by \citet{Wang_etal_2007} between the \ion{Si}{\sc II} $\lambda$6355 polarization five days before $B$-band maximum and $\Delta {\rm m}_{15}$, nor do they match the relation between the velocity gradient of \ion{Si}{\sc II} $\lambda$6355 and the associated peak polarization \citep{Maund_etal_2010}.  The mismatch may be due to SNe\,1999by and 2005ke perhaps being typical representatives of underluminous Type Ia SNe. Their spectroscopic and polarimetric properties can be understood within the frameworks of delayed detonations originating from a rapidly rotating WD or WD-WD mergers \citep{Patat_etal_2012}.

\section{Conclusions}
\label{sec_conclusions}
At five epochs between days $-$14.5 and $+$14.5 from maximum light, we have obtained high-quality optical VLT spectropolarimetry of the normal Type Ia SN\,2019np.  The first epoch of our observation is the earliest such measurement carried out to date for any Type Ia SN.  The data have been analysed with detailed radiation-hydrodynamic non-LTE simulations in the framework of an off-center delayed detonation which produces aspherical distributions in the burning products and, in particular, an aspherical $^{56}$Ni core.  The observations are also compatible with the presence of a central energy source that deviates from spherical symmetry. 
The understanding of SN\,2019np that we have achieved with our simulations can be summarised as follows.

\renewcommand{\labelenumi}{(\arabic{enumi})}
\begin{enumerate}
\item 
A viewing angle of $\sim 45^{\circ}$ provides the best fit to the amplitude and temporal evolution of the polarization spectra including \ion{Si}{\sc II} $\lambda$6355 and the continuum (Section~\ref{sec_model}). As discussed in \textcolor{black}{point} (3) below, the continuum polarization at the first epoch on day $-$14.5 requires a separate component. 

\item
The five epochs roughly cover the time interval in which the photosphere receded through the layers of incomplete carbon burning, complete carbon burning, incomplete and complete (QSE\footnote{QSE: Quasi-Statistical -Equilibrium}) explosive oxygen burning, and incomplete Si burning at the interface to NSE.  The cadence of the observations corresponds to a resolution of $\sim 6000$\,km\,s$^{-1}$ in expansion velocity.  Higher-cadence observations than obtained for SN\,2019np are required to resolve any structures in the ejecta at smaller scales.  For instance, considering the stratification in expansion velocity of the abundance layers and the recession speed of the photosphere, a cadence of $\sim 1$ day would be essential to map out the interfaces between different chemical layers. 

\item
The outermost $\lesssim 4 \times 10^{-3}$\,M$_{\odot}$ region seen during the first $\sim 3.5$ days after the SN explosion is consistent with a C/O-rich layer.  The change of the polarization position angle (Figure~\ref{Fig_qu}) and the rotation of the dominant axis (Figure~\ref{Fig_polar}) from the first to later epochs suggest a different orientation of the outermost layer compared to the inner regions.  To account for the outer asymmetry, a separate structure had to be added to the models.  This renders it relatively unlikely that SN\,2019np originated from a sub-$M_{\rm Ch}$ double detonation induced by a helium shell at the surface of the WD because the initial shape of the core of the WD can be expected to be symmetric, \textcolor{black}{being} governed by gravity.  
A deformation of the core may occur in a rapidly, differentially rotating WD with extreme specific angular momentum close to the center \citep{Eriguchi_Mueller_1993}, but is not likely. 

\item
Although the continuum polarization on day $-$14.5 was only 0.21\%$\pm$0.09\%, the hydrodynamic modeling of the polarization with an evolving density profile suggests the presence of a remarkably aspherical outermost layer of the SN, comprising a fraction of $\sim$ (2--3) $\times 10^{-3}$ of the WD-progenitor mass.  In conjunction with the inclination of $\sim 45^{\circ}$, the axis ratio of the electron density distribution predicted by the models amounts to $\sim 2$--3.  That is, in the presence of a steep density gradient in the outermost layers, a low but nonzero continuum polarization cannot be taken as an indicator of a high level of sphericity. 

\item
The rise of the continuum polarization to 0.19\%$\pm$0.07\% about two weeks after peak luminosity is consistent with an aspherical $^{56}$Ni distribution in the core as predicted by off-center delayed-detonation models. From the models, a change of the polarization position angle by $\sim 20^{\circ}$ can be expected.  The time-evolving distribution of the polarimetric signals in the polar plots (Figure~\ref{Fig_polar}) and the rotating dominant axis on the $Q$--$U$ plane (Figure~\ref{Fig_qu} and Section~\ref{sec_quplane}) are consistent with this prediction.  Around maximum light, the off-center contribution causes a continuum polarization of $\lesssim 0.1$\%.

\item
\textcolor{black}{Small $p^{\rm cont}$ seems to be a characteristic of off-center DDT models. This justifies the assumption of negligible intrinsic polarization made for the determination of the ISP (Section \ref{sec_isp}).  This method differs from those commonly applied to core-collapse SNe (Section \ref{sec_isp}). The difference between the continuum polarization $p^{\rm cont}$ and the average polarization over the entire spectrum with many line-dominated spectral regions remains small, $\sim 0.1$\% from $-6.4$ to +0.5 days for our normally bright off-center DDT model. The difference is consistent with the observations of SN\,2019np. \textcolor{black}{To understand} the physical reason for being small, see  Section \ref{Sec_pol_spec}. Despite much stronger line polarization, a similar small value of $p^{\rm cont}$ and a small difference in the continuum is found for the Type Ia SN\,2004dt when analysed within the framework of off-center DDT models (Section \ref{sec_other_sne}).}

\item
The polarization observed across \ion{Si}{\sc II} $\lambda$6355 on day $-$14.5 is higher than in our model (Figure~\ref{Fig_pol_model}). The dominant axes fitted to this line and the optical continuum are both not well defined at this phase (Sect.~\ref{sec_quplane}). 
From our full-star models, it is hard to deduce whether the asphericity of these two components in the outermost layers has a common origin --- for instance, owing to interaction with a low-mass accretion disc, which is a free parameter and remains unconstrained in the current modeling process of the full star.  However, the toy model for the continuum polarization (Section~\ref{sec_contpol}) demands different symmetry axes for the density and the abundances.

\item
High asphericity is found to be confined to the very outer layers.  It emerges from the fitting when introduced as an additional free parameter not included in the hydrodynamical model.  Possible physical causes may include a short-lived interaction between the SN ejecta with a low-mass accretion disc \citep{Gerardy_etal_2007} or a companion star \citep{2000ApJS..128..615M}, or surface burning as found in sub-$M_{\rm Ch}$ explosions \citep{Shen_etal_2012} but see item (3), or the imprint of the burning of H/He-rich material originating from the surface \citep{Hoeflich_etal_2019}.  Interaction with a donor star seems less likely because it would affect not only the outermost layers.  As shown in Figures~\ref{Fig_iqu_ep1} to \ref{Fig_iqu_ep3} and discussed in Section~\ref{sec_model}, the sub-$M_{\rm Ch}$ model is in tension with the observation of the \ion{Ca}{\sc II}\,NIR3 feature.  A similar early-time polarization has been observed in the normal-bright SN\,2018gv \citep{Yang_etal_2020}. However, in the underluminous SN\,2005ke, a significant polarization was observed in the outer $\sim 0.2$\,M$_{\odot}$, hinting toward rapid rotation or a dynamical merger \citep{Patat_etal_2012}.  

\item
The continuum polarization of SN\,2019np vanished by day $-$11.5 and {remained consistent with zero within one $\sigma$} until the SN reached its peak luminosity, indicating a high degree of spherical symmetry between the outermost 0.02--0.03\,M$_{\rm WD}$ and $\sim 0.5$\,M$_{\rm WD}$.  In the case of a highly aspherical configuration extending into deeper layers, the continuum polarization should increase with time since the density distribution flattens significantly and the scattering optical depth decreases from the optically thick regime.  However, such an increase in continuum polarization was not observed in SN\,2019np.  This partly invalidates the conventional assumption that a low continuum polarization provides evidence of low asphericity.  In the presence of a steep density gradient as in the outermost layers of SN\,2019np, major deviations from spherical symmetry are well possible. \textcolor{black}{High-cadence observations are needed to distinguish between these alternatives.}

\item 
The increased continuum polarization on day $+$14.5 can be explained by abundance asphericities in an off-center delayed detonation.  The direction of the dominant axis of SN\,2019np has also changed between days $+$0 and $+$14.5 (Figure~\ref{Fig_qu}), which may indicate a small, off-center distribution of the central energy source, 5\%--10\% of the total amount of $^{56}$Ni.  The model also predicts a change of the polarization position angle of the quasi-continuum.  Although qualitatively indicated by the observations, the change in the polarization position angle of the continuum is hard to quantify from the available observations because the intrinsic continuum polarization is very low (Section~\ref{sec_spec_pol}), and the numerous small wiggles at this low level may also cause a problem (Figures~\ref{Fig_tau_max}--\ref{Fig_pol_model}).

\item
None of the possible mechanisms causing the rise of the \ion{Ca}{\sc II}\,NIR3 polarization on day $+$14.5 has been included in our simulations.  Its origin remains uncertain, as discussed in Section~\ref{sec_pol_spec}.

\item
In the optical domain, spectral line formation and polarization by electron scattering take place in the same region of the expanding atmosphere (Section~\ref{sec_pol_spec} and Figure~\ref{Fig_tau_max2}).  By contrast, the canonical polarization-by-obscuration picture \citep{Wang_wheeler_2008} requires that spectral lines are formed mainly above the last continuum-scattering surface.
{ To first order, this approximation can be used to place an upper limit on the peak polarization from large-scale asymmetries, which are produced by the large Sobolev optical depth over the entire photosphere, if the formation of the quasi-continuum is dominated by Thomson scattering. In SNe~Ia, the \ion{Si}{II}\,$\lambda\lambda$6348, 6373 doublet provides an example.  It originates from a low-excitation state, and Si accounts for $\sim$60\% of the total mass fraction corresponding to up to $\sim 2,500$ times the solar value found in the H-rich envelopes of CC-SNe.  The expansion velocities of Si-rich layers range from $\sim 9000$ to more than 22,000~km~s$^{-1}$.
%The other the Ca II is a resonance line even for solar abundance. 
%\textcolor{blue}{\bf This previous sentence is muell.}
In the presence of large-scale asphericity in Si, this line will be significantly polarized unless the region of asymmetric density or Si abundance is hidden behind an extended photospheric region.  
%\textcolor{blue}{\bf (DBA:  Added "roughly spherical". 
%\textcolor{red}{PAH: "roughly spherical" is not a requirement to see no line polarzation in Si. An aspherical structure may increase the continuum polarization or that of other lines.} The rest of the sentence is muell.)}\textcolor{red}{PAH: rest delected.}
%and neglecting of multi-scattering causing $p$ to be an upper limit. 
For small-scale or multiple structures, the resulting polarization depends sensitively on $\tau_{\rm sc}$ at their location. %\textcolor{red}{of the small scale structures.}  
%\textcolor{blue}{\bf (DBA:  Which location?  Location of what?)}\textcolor{red}{PAH: added small scale structures, but is it needed?}
%leading to the loss of a unique relation.}
%\textcolor{blue}{\bf DBA:  Relation between what quantities?} \textcolor{red} {redundant and deleted.}
}

\item
Overall, the polarization spectra and their temporal evolution can be understood as a variable thermalisation optical depth and partial blocking of the photosphere at a given geometric depth. 

\item
There are strong high-amplitude fluctuations in the polarization spectra 
%\textcolor{blue}{\bf \ion{Ca}{\sc II}\,NIR3}
on day $-$14.5.  They can be attributed to the strongest S/Si and the \ion{Ca}{\sc II} lines, in particular { \ion{Ca}{\sc II}\,NIR3}.  The lack of similar patterns and the agreement between the observations and synthetic spectra in the spectral region dominated by iron-group elements, namely the $U$ and $B$ bands and longward of 8500\,\AA, seem to rule out Ni, Co, or Fe \textcolor{black}{as being affected by the same mechanism(s) responsible for the fluctuations just mentioned.}.  For details and their possible origin see Section~\ref{Sec_pol_spec}.

\item 
At the same time, there are also flocculent structures in the polar diagram 
{ for \ion{Ca}{\sc II}\,NIR3}
(Section~\ref{sec_line_pol}, Figure~\ref{Fig_pol_model}) with a broader distribution in position angle and a lower polarization degree (see Figure~\ref{Fig_polar}), suggesting a more complex structure of the outermost layers than adopted in the models.

\item
Many relatively weak polarization features within the wavelength range $\sim 4500$--6000\,\AA\ and beyond 6800\,\AA\ can be interpreted as signatures of spectral line blends { and unresolved wiggles}.  The general agreement between data and model suggests that many of these weak polarization features are real.  Some discrepancies do not necessarily invalidate this conclusion, but likely point toward small-scale structures (Section~\ref{sec_pol_spec}).  However, the noise in the data for SN\,2019np sets a limit to probing those scales in this event.

\item
Polarization is able to pick up spectral signatures not visible in the flux spectra because lines depolarize but both absorb and emit photons (see Section~\ref{Sec_pol_spec}). Spectropolarimetry of sufficient spectral resolution can reveal spectral lines that are undetectable in flux spectra, and, at proper cadence, their depth of formation can be inferred.  This added diagnostic power is independent of any asphericity and can be important to discriminate explosion scenarios.

\item
The asphericity in the $^{56}$Ni distribution is significant (Figure~\ref{Fig_structure}) although the continuum polarization is relatively low.  As discussed in Section~\ref{sec_contpol}, a low thermalisation depth results in vanishingly low continuum polarization, except for photons that graze the photosphere at low optical Thomson-scattering depth (see Section~\ref{Sec_pol_spec}).  
As a corollary, high polarization can be expected if the off-center component dominates, but polarization may fail to detect even significant asphericity in the density produced by a Fe/Co/Ni core. 

\item
The polarization profiles of strong isolated features provide new diagnostics to probe for mixing on small and medium scales, and to explore the chemical stratification (Section~\ref{sec_pol_spec}).

\end{enumerate} 

%\textcolor{red}{\noindent{\sl Brief summary of off-center DDT models:} 
%}

\section{Future Opportunities}~\label{sec_summary}

{
\subsection{The Diagnostic Power of Spectropolarimetry of Type Ia SNe}
\label{sec_opportunities}
}

Asphericity holds a key to the understanding of the nature of thermonuclear SNe.  It involves three main components: (1) the exploding WD, (2) all other matter bound in the progenitor system, which may include a companion star and any bound CSM such as a common envelope or an accretion disc, and the inner parts of several winds (e.g., from the WD, a companion star or its Roche lobe, and an accretion disc), and (3) the unbound CSM consisting of the outer faster and less dense parts of the winds and ultimately the interstellar medium. 

A detailed discussion of the effects of these three constituents on the geometry is beyond the scope of this paper.  A broad introduction to the geometrical signatures in polarization and nebular spectra expected from all commonly considered explosion models and progenitor channels was recently given by \citet {2021ApJ...922..186H}.  Reviews on various topics can be found among the articles collected in \citet{2017hsn..book.....A}.

\begin{figure}
\includegraphics[width=1.0\linewidth]{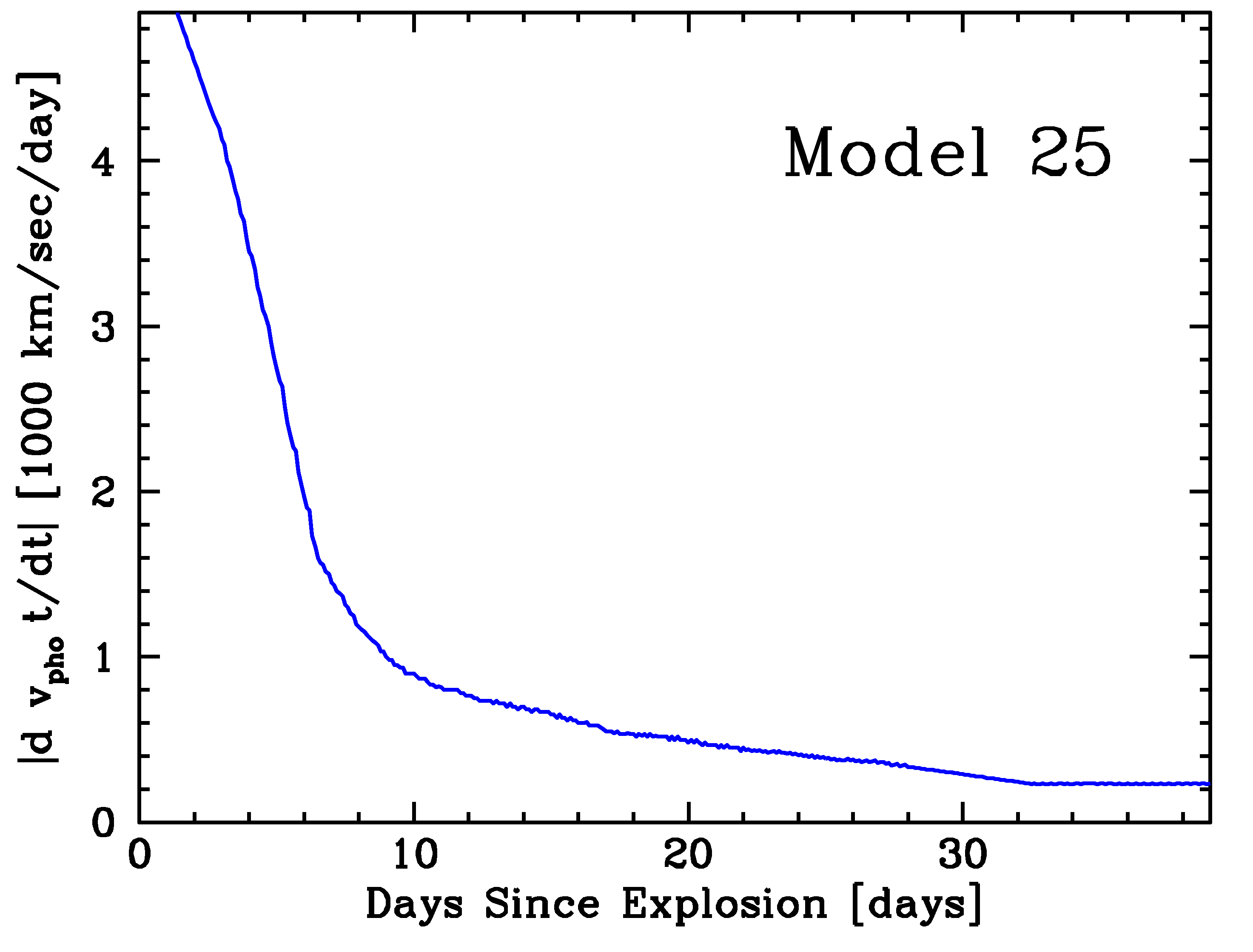}
\caption{
Rate of recession of the photosphere in the spherical high-resolution Model~25 as a function of time.  The recession rate $d v_{\rm phot}/dt$ for Model 25 (see Section~\ref{sec_model}) was calculated using the Rosseland mean opacity in the optical and a range of $\pm2$ days to determine $v_{\rm phot}$.  The exact values depend on the explosion model used (see Figure~9 of \citealp{Quimby_etal_2007}). 
\label{Fig_rec}
}
\end{figure}

\renewcommand{\labelenumi}{\alph{enumi})}
\begin{enumerate}
\item
To properly plan observing sequences, timescales are critical.  For all explosion paths, the initial phase of the explosion takes a few seconds to a minute, and the main spatial dimension is given by the exploding WD or the two merging WDs and ranges from $1.5 \textcolor{black}{\times} 10^8$ to $10^9$\,cm.  The hydrodynamical interaction of the explosively expanding envelope with a companion star, any accretion disc, and the inner parts of the wind(s) takes place within $\sim 10^{10}$--$10^{13}$\,cm.  Considering the velocities and masses in the outer layers of the explosion, this corresponds to timescales of minutes to about an hour for interaction with the bound matter.  Interaction with the wind(s) and the ISM can extend over days to many years, and is only limited by the transition to the supernova-remnant phase. 

\item
The impact of the various components on the geometrical structure and the associated polarization depends on the mass of the companion and any bound CSM relative to that of the ejecta which differs strongly between the explosion processes ($\sim 0.6$--2\,M$_{\odot}$).

\item
The mass-loss rate from the system may range from $\sim10^{-6}$ to $10^{-4}$\,M$_{\odot}$\,yr$^{-1}$ and can be due to any model-dependent combination of the wind from the WD, the wind from a companion,  
super-Eddington accretion onto the WD (e.g., \citealp{Nomoto_etal_1976}), the Roche lobe in a single-degenerate system, or the high-velocity wind from accretion discs in cataclysmic variables, or it may be produced during the final phase of dynamical mergers.  
Upper limits to the total mass content of these winds integrated over the time considered for early polarimetry are $\sim10^{-4}$--$10^{-7}$\,M$_{\odot}$ \citep{Dragulin_etal_2016,2017hsn..book..875C}.  Even at the earliest times when significant polarization can still be expected, the dynamical effects of the impact of the ejecta on the unbound CSM are likely to be small (Figure~\ref{Fig_structure}).  Nevertheless, in early-time photometry, some small additional blue flux may appear owing to energy released during the interaction.  For very high wind densities, the hard radiation at the shock discontinuity and the reverse shock may lead to enhanced ionisation in the photosphere of the SN.  In SN\,2019np, we see no obvious evidence of such effects (Section~\ref{sec_pol_spec}).

\item
The CSM bound in the system may include matter in a Roche lobe and/or an accretion disc.  Observational evidence has been reported by (for example) \citet{2006ApJ...650..510A}, and, from high-velocity \ion{Ca}{\sc II} absorptions in early-phase spectra, its mass has been estimated to be of order $10^{-2}$--$10^{-3}$\,M$_\odot$ \citep{Gerardy_etal_2007}.  This mass is comparable to that in the layers probed by our early polarimetry of SN\,2019np, and it is compatible with the large asymmetries proposed (Figure~\ref{Fig_structure}).  The small-scale structures depend on the scale height of the material (e.g., the Roche lobe or the disc) and the sound-crossing time during the hydrodynamical phase of the interaction.  However, since the mass of the bound CSM is much larger than that contained in the wind, the structure produced by the interaction of the ejecta with this matter can be expected to be conserved in the subsequent possible interaction with the outer winds and the ISM, although the temperatures and sound speed are likely to be high \citep{Margutti_etal_2014,2020ApJ...900..140H}.  Therefore, polarimetry will provide a unique way to explore the bound CSM.  As discussed above (Sections~\ref{sec_quplane} \& \ref{Sec_pol_spec}), \textcolor{black} {some of } the wiggles in polarization spectra and flocculent structures in polar plots seen in the spectropolarimetry of SN\,2019np may have their origin in Rayleigh-Taylor or Kelvin-Helmholtz instabilities and crossing shock waves produced during the injection with an orientation imprinted by the bound CSM.  Polarization measurements with a latency between hours and one day are needed (Figure~\ref{Fig_rec}) to learn whether the polarization position angle persists, which would suggest a large-scale structure, dominated by instabilities imprinting their characteristic size as wiggles and large-amplitude fluctuations in the polarization spectra ans well as flocculent structures in the polar diagrams, or a combination of large and small scales.
These structures become most prominent in  \ion{Ca}{\sc II}\,NIR3, which is an excellent tracer of structure \textcolor{black}{owing} to its large atomic cross section.

\item
Alternatively, the early wiggles and the flocculent structures in SN\,2019np may be related to the explosion mechanism, in which case they would have an origin internal to the SN proper.  Potential sources are explosive surface He-burning in sub-$M_{\rm Ch}$ explosions (e.g., \citealt{Shen_Moore_2014}) or in NSE-rich material rising from the central region in gravitationally confined thermonuclear explosions \citep{2005ApJ...622L..41K}.  In both cases, some 0.02--0.1\,M$_\odot$ of $^{56}$Ni may reach near-surface regions.  This is well within the mass range that can be probed by polarization as in SN\,2019np (Figure~\ref{Fig_structure}) because the characteristics and especially the structures associated with these processes are different.  A central distinguishing criterion is the presence of products from low-density burning in sub-$M_{\rm Ch}$ explosions and of NSE-dominated material from high-density burning and a mixture of NSE and QSE in gravitationally confined detonations.

Prompted by detections of early excess luminosity in optical light curves, \citet{Piro_etal_2018} suggested $^{56}$Ni as a possible energy source.  At a first glance, early photometry of iPTF16abc \citep{Miller_etal_2018} and SN\,2019np \citep{Sai_etal_2022} makes it plausible that both light curves can be well explained by $^{56}$Ni mixed into the outer layers of the ejecta.  A more detailed study has been presented by \citet{2022MNRAS.642.189M}.  The authors proposed that the bump in the light curve can be understood by a Ni shell of 0.02--0.03\,M$_\odot$ in the outer 0.2\,M$_\odot$ but also noted that the colours are too blue, and the spectra would be dominated by Fe/Co/Ni.  Since both of these side effects are not supported by the observations, the authors suggested as a remedy that the inferred 10\% of $^{56}$Ni is concentrated in a small clump.   

However, the observed flux and polarization spectra safely rule out $^{56}$Ni-powered early-time light curves (Section~\ref{sec_pol_spec}).  Moreover, the proposed small clump is unlikely to be the \textcolor{black}{correct explanation;} the luminosity would be dominated by a plume resulting in a flip in polarization position angle (Figure~\ref{Fig_pol_axis_ratio}) because even as little \textcolor{black}{as} 0.02\,M$_\odot$ of $^{56}$Ni would dominate the energy input.  In both the sub-$M_{\rm Ch}$ and the gravitationally confined detonation, the hypothetical $^{56}$Ni would be in the surface layer. It would contribute 50\% of the total heating and be equivalent to the energy output from 0.1\,M$_\odot$ of $^{56}$Co, and may not solve the blue-colour problem.  The latter may be an opacity effect in a rapidly expanding atmosphere instead of a heat indicator \citep{2022ApJ...932L...2A}.  The presence of burned material from a He-triggered sub-$M_{\rm Ch}$ explosion can be clearly demonstrable with spectropolarimetry because it can detect iron-group elements down to solar metallicity (Section~\ref{sec_pol_spec}), which is not possible from \textcolor{black}{total-}flux spectra alone.  Furthermore, polarimetry is the ideal tool to distinguish explosions without surface burning and He-triggers in sub-$M_{\rm Ch}$ WDs. 

\item
Interaction of the SN ejecta with a companion star is a consequence of most explosion models except for the violent, dynamical, and secular mergers.  The early bump identified in the photometry of SN\,2017cbv appears blue and has been modeled by interactions with a subgiant star at a distance of 56 solar radii from the exploding WD \citep{Hosseinzadeh_etal_2017}.  However, the model also predicts a stronger \textcolor{black}{ultraviolet} flux than was observed.  All SN mechanisms with an internally triggered explosion seem to have a problem explaining the early blue excess flux and for the explanation \textcolor{black}{to} resort to interaction with some CSM.  An interaction with a relatively high-mass object can be expected for many progenitor systems, and it would leave its imprint not only in the surface layer but all the way down to the central region.  So large a structure is likely to become visible in polarization.  But it may also produce small-scale structures in the abundances and, depending on the donor star, the density \citep{2000ApJS..128..615M}.  Interaction with a companion causes a tight connection between outer and inner regions of the expanding envelope, including a common and persistent symmetry axis and, initially, a cone with an opening angle of $\sim 30^{\circ}$.  Although our modeling of SN\,2019np did not include a companion, the observed change in polarization angle (Figures~\ref{Fig_iqu_ep1}--\ref{Fig_iqu_ep4}) probably disfavours a dominant effect of an ejecta/companion interaction on the polarization (Figure~\ref{Fig_pol_model}).  Any such effect depends on the size of, and distance to, the companion star, and the large-scale asymmetry imposed by a small companion can be expected to be largest in deep layers \citep{2000ApJS..128..615M}.  Furthermore,  in contrast to an off-center DDT, a small-scale structure produced by Rayleigh-Taylor instabilities will occupy a cone instead of a spherical layer.  Unfortunately, in our study of SN\,2019np, we lack the cadence to resolve small scales and their distribution from the outer to the inner regions.
%\textcolor{black}{\bf PAH: ok but black}
%\textcolor{blue}{\bf DBA:  Black edits.}

\item
For deflagration fronts, we must expect small-scale Rayleigh-Taylor instabilities and the imprint of the thermonuclear runaway, the caustic distribution of the burning processes, and magnetic fields as discussed throughout this paper (especially Sections~\ref{sec_pol_spec} and \ref{sec_conclusions}) and by \citet{2021ApJ...923..210H}.  Polarimetry of \ion{Ca}{\sc II}\,NIR3 is a highly sensitive tool to probe for the associated structures.  The polarization of this triplet in the central region is strong evidence for mixing from the outside because burning to hot NSE destroys Ca (Figure~\ref{Fig_structure}).  However, the huge atomic cross section of \ion{Ca}{\sc II}\,NIR3 desensitises it to abundance effects (Section~\ref{sec_pol_spec}).  When observations of this feature and the continuum polarization with similar characteristics as ours of SN2019\,np but with higher cadence are combined with high-resolution nebular spectra in the \textcolor{black}{near- and mid-infrared} at later epochs, a fairly complete \textcolor{black}{three-dimensional} model of the structure of thermonuclear SNe can be assembled \citep{Kotak_etal_2004,Telesco_etal_2015,2021ApJ...922..186H}.

\item
Off-center DDT explosions of $M_{\rm Ch}$ WDs can generally be thoroughly investigated by spectropolarimetry.  For sub-$M_{\rm Ch}$ explosions, the location of the secondary ignition of the C/O core can only be expected to show an imprint on the polarization spectrum if it happens in the regime of distributed  burning.

\item
For SN\,2019np, we found evidence of an off-center component in the $^{56}$Ni distribution, which disfavours sub-$M_{\rm Ch}$ explosions.  However, we also showed that large-scale asymmetries in the extended central Fe/Co/Ni region cannot be excluded (Section~\ref{sec_contpol}). Contrary to normal-bright SNe and SN\,2019np in particular, polarimetry of underluminous Type Ia SNe provided strong evidence for significant asphericity even in deeper layers.  This may favour fast rotating WDs or dynamical mergers (Section~\ref{sec_conclusions}). 
A possible discriminator may be the presence of Rayleigh-Taylor instabilities in models with deflagration burning, and their absence in pure detonation models like dynamical mergers \citep{2017hsn..book.1237G}.  As explained above, time-resolved polarimetry can establish the actual facts.

\item
The Introduction has discussed polarization of Type Ia SNe as a diagnostic tool, also in connection with minority events like violent mergers \citep{2017hsn..book.1257P,Kushnir_etal_2013}.  Such events should exhibit prominent polarization signatures because the off-center component would dominate (Figure~\ref{Fig_pol_axis_ratio}).  Therefore, the amplitude in the continuum polarization with time would be larger by a factor of 5--6, and large polarization can also be expected in spectral lines.

\end{enumerate}

\subsection{Implications for the Design of Spectropolarimetric Observing Sequences and Simulations}
\label{Sec_planning}

Spectropolarimetry with adequate cadence from the earliest possible moment and (to potentially expose the EC layers) up to $\sim 3$ weeks after maximum \textcolor{black}{brightness} is necessary to get a comprehensive picture of normal Type Ia SNe on all scales and to discriminate between competing interpretations. 
The \textcolor{black}{SNR} must be high, and the spectral resolution should be better than that of our SN\,2019np data.  In our observations of SN\,2019np, the combination of a spectral resolving power of $R \approx 440$ and a $\sim 4$--6 day cadence only achieved a resolution of $\sim 6000$\,km\,s$^{-1}$ in comoving-frame velocity between mass elements.  
This limited our analysis of any smaller-scale structure in the SN ejecta and the surface layer.  Observations with low cadence disqualify for several quantitative comparisons with models, most importantly the identification of individual clumps (predicted by Rayleigh-Taylor instabilities), layers of specific chemical elements, the interface between the Ca-rich and the inner NSE-region (Figure~\ref{Fig_structure}), and the origin of the asphericity in the outermost layers.

To devise a more optimised observing strategy, the evolution of the recession velocity of the photosphere needs to be considered, which Figure~\ref{Fig_rec} illustrates for Model~25.  Starting $\sim 1$ week after explosion, during the rising and photospheric phases of Type Ia SNe, this velocity declines from $\sim 1000$ to $\sim 200$\,km\,s$^{-1}$\,d$^{-1}$. 
Accordingly, a 2--4 day observing cadence suffices to locate the interfaces between various chemical constituents relevant for the internal structure of the explosion (items (e)--(j) above and Fig. \ref{Fig_structure}).  With one-day cadence and a matching spectral resolving power of 1000 (corresponding to a transversal resolution of 300\,km\,s$^{-1}$), the morphology and origin of the individual structures discussed in items (e)--(j) can also be explored, { namely the properties of the burning front including secondary detonations or a DDT, shear instabilities due to interactions with a companion star, the imprint of the thermonuclear runaway, the consequence of mixing of radioactive and EC elements, and possibly magnetic fields.}

To also determine the shape of the outermost layers of the exploding WD calls for rapid-response and high-cadence observations because the radial density index governing this shape changes early and rapidly.  Some 3.5 days after the explosion, a one-day cadence corresponds to a radial resolution of $\sim 3500$\,km\,s$^{-1}$ and $10^{-4}$\,M$_{\odot}$ in mass (Figure~\ref{Fig_structure}). Therefore, such observations resolve both the bound CSM (item (d) above) and the He layer and $^{56}$Ni mass (item (e)) to be expected for sub-$M_{\rm Ch}$ or confined detonation models.  At an expansion velocity of 24,000\,km\,s$^{-1}$ (a typical value of high-velocity components in \ion{Ca}{\sc ii}\,NIR3 and the cutoff velocity of the blue wings in \ion{Si}{\sc ii} $\lambda$6355; \citealp{Quimby_etal_2007}; \citealp{Gerardy_etal_2007}), the transversal resolution 
{(which is set by the spectral resolution of 300\,km\,s$^{-1}$)}
exceeds the radial resolution by a factor $\sim 10$. 
{ These resolutions describe cones with approximate opening angles of $20^{\circ}$ and $2^{\circ}$ for the radial and transversal directions (respectively), and are respectively governed by the cadence and the spectral resolution.\footnote{ The opening angles are given by the arcsin of the structure size divided by the photospheric radius in velocity space. }
The apices of the cones are at the center of the WD and the numerical grid.  The opening angles are comparable to the angle suspended by (for example) the region affected by an interaction with a companion star, or about 20\% and 2\% of the photospheric radius. }

With such data, the interaction of the ejecta with an accretion disc, a Roche lobe (item (d) above) and a progenitor star (item (f)) can be significantly distinguished.  They may also be sufficient to characterise the structures and underlying physical mechanism(s) which produce the large-amplitude fluctuations in the polarization profile and flocculent features in polar diagrams of strong spectral lines.

The numerical models applied in this study have an {effective spatial resolution of $\sim 600$\,km\,s$^{-1}$ for clumps.\footnote{The effective resolution is estimated from the grid resolution, i.e., the domain size ($2 \times 25,000$\,km\,s$^{-1}$) divided by the number of grid points (330) in a differential scheme (contributing a factor of $\sim 2$).  For spherical clumps as the simplest structures, it is multiplied by another factor 2. This results in 75\,km\,s$^{-1} \times 2^3$.

}
This is just sufficient to resolve the largest Rayleigh-Taylor instabilities, for example, but already appears to be higher than all SN spectropolarimetry obtained to date.}
Higher-resolution models to resolve small-scale structures can be achieved with sub-star instead of full-star simulations or, better, a space-domain implementation\footnote{As part of an ongoing PhD project, the domain deposition is currently being implemented using ``pancake slicing''  and cycling between the three Cartesian coordinates.}  for our Variable Eddington Tensor solver in HYDRA as discussed  by \citet[][and references therein]{2021ApJ...923..210H}.  Investigation of the effects caused by details of the thermonuclear runaway in a $M_{\rm Ch}$ explosion or any secondary C/O ignition would require spectropolarimetry at late times in combination with late-time near- and mid-infrared flux spectra \citep{2021ApJ...922..186H}.  Spectropolarimetry will provide a tomographic sampling of the geometric properties of the layer near the Si/Fe interface in the inner regions of the WD remnant, while nebular flux profiles at sufficient spectral resolution will probe the physical conditions and kinematics of this region through unblended line profiles, which are also sensitive to the aspect angle of the observer \citep{2021ApJ...922..186H}.  The combination of these two datasets will subject models to critical consistency checks because they investigate the same region from different perspectives.  Analogous simulations should be employed to test other models and make use of hydro-dynamical simulations to investigate points (e)--(j) in the list above. \textcolor{black}{A detailed discussion of many other scenarios is beyond the scope of this paper. Obviously, similar simulations can  be applied to a wide variety of explosion scenarios and other transients \citep[e.g.,][]{2012AIPC.1429..204L,2019MNRAS.482.4057M,2021A&A...651A..19D,2021MNRAS.506.4621B}, and those will be performed in the future.}

\textcolor{black}{By using details previously not considered in combination with \textcolor{black}{extensive} modelling, we obtained new insights
into the formation of $p$ and obtained many results for SN\,2019np in Section \ref{sec_conclusions}. However, we pushed the analysis to the limit of current data and modelling efforts. This study should be regarded as a pathfinder for a new approach to the analysis of SNe~Ia data, to evaluate the limits and to identify the potential and shortcomings of current observations and theoretical models, to evaluate methods to correct for the ISM, and to develop future  polarization programs (Section \ref{sec_summary}).}

\bigskip
\noindent
{\sl Acknowledgements:}
We are grateful to the European Organisation for Astronomical Research in the Southern Hemisphere (ESO) for the generous allocation of observing time.  
We especially thank the staff at Paranal for their proficient and highly motivated support of this project in service mode. 
P.H.\ acknowledges support by the National Science Foundation (NSF) through grant AST-1715133.
\textcolor{black}{A.V.F.'s supernova group at U.C. Berkeley is grateful for financial assistance from the Christopher R. Redlich Fund and many individual donors, including Gary and Cynthia Bengier, Clark and Sharon Winslow, Sanford Robertson, Sunil Nagaraj, Landon Noll, and Sandy Ottelini. The research of Y.Y.\ has also been supported through a Benoziyo Prize Postdoctoral Fellowship.}

\bigskip
\noindent
{\sl Facilities:}
The observations were obtained with FORS2 and the Very Large Telescope at the European Southern Observatory's La Silla Paranal Observatory in Chile.  The simulations have been performed on the computer cluster of the astro-group at Florida State University. 

\bigskip
\noindent
{\sl Software:}
IRAF is distributed by the National Optical Astronomy Observatories, which are operated by the Association of Universities for Research in Astronomy, Inc., under cooperative agreement with the National Science Foundation.  HYDRA and various modules.  OpenDx, an open-source graphics package by IBM.
\bigskip
\noindent

\noindent
{\sl Data Avaibility Statement:} All data are available on request.

\begin{table*}
\begin{center}
\caption{Log of spectropolarimetry of SN\,2019np~\label{Table_log_specpol}.}
\begin{small}
\begin{tabular}{cccccccc}
\hline
\hline 
Epoch &  Object     &  MJD       &  Date             & Phase$^a$ &  Exposure    & Grism /  &  Airmass   \\
      &                &       &  (UT)    &  (day)     &   (s)                    & Resol.\ Power & Range \\
\hline
1$^b$ & SN\,2019np  & 58495.270  &  2019-01-12 06:30 &   $-$14.5 & 4$\times$550 &     300V/440     &  1.84--1.73   \\
      &             & 58495.298  &  2019-01-12 07:10 &           & 4$\times$550 &     300V/440     &  1.73--1.70   \\
\hline
2     & SN\,2019np  & 58498.302  &  2019-01-15 07:14 &   $-$11.4 & 4$\times$600 &     300V/440     &  1.71--1.70   \\
\hline
3     & SN\,2019np  & 58503.305  &  2019-01-20 07:18 &    $-$6.4 & 4$\times$210 &     300V/440     &  1.70--1.71   \\
      & HD\,93621$^c$ & 58503.322  &  2019-01-20 07:43 & --        & 1$\times$0.51  &     300V/440     &  2.13   \\
\hline
4     & SN\,2019np  & 58510.272  &  2019-01-27 06:31 &   $+$0.5  & 4$\times$360 &     300V/440     &  1.71--1.70   \\
\hline
5     & SN\,2019np  & 58524.188  &  2019-02-10 04:31 &  $+$14.5  & 4$\times$360 &     300V/440     &  1.71--1.70   \\
\hline
\end{tabular}\\
{$^a$}{{ Relative to the estimated peak on MJD 58509.7 / UT 2019-01-26.7; MJD and Date are given as the start time of the CCD exposure.}  } \\
{$^b$}{Epoch 1 observation consists of two sets of exposures at four half-wave plate angles.} \\
{$^c$Flux standard, observed at a half-wave plate angle of $0^\circ$.}
\\
\end{small}
\end{center}
\end{table*}

\begin{table*}
\caption{%Spectropolarimetic properties of SN\,2019np. 
Values of least-squares fitting parameters on the $Q$--$U$ plane for SN\,2019np
\label{Table_pol}}
\begin{scriptsize}
\begin{tabular}{c|cc|cc|cc|cc}
\hline
\hline
\#Epoch   & $Q^{\rm cont}$    & $p^{\rm cont}$  &  $\alpha$ / $\alpha^*$      &  $\theta_{d}$ / $\theta_{d}^*$          &  $\alpha^{\rm Si\,II\lambda6355}$   &  $\theta_{d}^{\rm Si\,II\lambda6355}$  &  $\alpha^{\rm Ca\,II\,NIR3}$   &  $\theta_{d}^{\rm Ca\,II\,NIR3}$ \\
Phase$^a$  & $U^{\rm cont}$    & [\%]            &  $\beta$ / $\beta^*$            &  (deg)                 &   $\beta^{\rm Si\,II\lambda6355}$   &  (deg)                                 &  $\beta^{\rm Ca\,II\,NIR3}$    &  (deg) \\
\hline
\# 1      &   0.194$\pm$0.125 & 0.209$\pm$0.087 & $-$0.061$\pm$0.142 / $+$0.478$\pm$0.131  &  23.8$_{-18.2}^{+7.9}$ / $-$33.5$_{-2.7}^{+4.8}$  &  0.49$\pm$0.13                          &  $-$31.4$_{-3.5}^{+6.6}$              &  $+$0.478$\pm$0.490              &  -32.0$_{-5.8}^{+26.1}$            \\
$-$14.5 d &   0.077$\pm$0.073 &                 &  1.093$\pm$0.895 / $-$2.347$\pm$      0.791   &                        &  $-$1.95$\pm$0.77                     &                                        &  $-$2.045$\pm$1.836                   &  \\
\hline
\# 2      &   0.008$\pm$0.155 & 0.099$\pm$0.080 & $-$0.105$\pm$0.005 / $-$0.108$\pm$0.004 &  2.1$_{-1.3}^{+1.3}$ / 1.4$_{-1.0}^{+1.0}$   &  $-$0.095$\pm$0.012                     &  5.6$_{-2.0}^{+1.9}$                   &  $-$0.153$\pm$0.015                &  $-$6.2$_{-2.3}^{+2.4}$          \\
$-$11.4 d &$-$0.098$\pm$0.078 &                 &  0.0745$\pm$0.046 / 0.050$\pm$0.034  &                        &   0.198$\pm$0.071                       &                                        &  $-$0.219$\pm$0.086                &  \\
\hline
\# 3      &$-$0.037$\pm$0.091 & 0.075$\pm$0.080 & $-$0.065$\pm$0.005 / $-$0.058$\pm$0.005 &  2.4$_{-1.6}^{+1.6}$ / 8.6$_{-1.4}^{+1.3}$  &  $-$0.026$\pm$0.018                   & 14.2$_{-2.2}^{+2.1}$                   &  $-$0.131$\pm$0.018                &  18.4$_{-3.4}^{+2.9}$            \\
$-$6.4 d  &$-$0.065$\pm$0.081 &                 &  0.085$\pm$0.055 / 0.310$\pm$0.051 &                        &  0.541$\pm$0.097                        &                                        &  0.748$\pm$0.171                   &  \\
\hline
\# 4      & 0.109$\pm$0.107   & 0.110$\pm$0.100 & $+$0.0056$\pm$0.0044 / $-$0.0026$\pm$0.0036 &  0.1$_{-1.2}^{+1.2}$ / 4.3$_{-1.0}^{+1.0}$  &  $-$0.013$\pm$0.011                   & 5.3$_{-1.9}^{+1.8}$                   &  $-$0.107$\pm$0.017              &  20.9$_{-3.2}^{+2.7}$            \\
$+$0.5 d  &$-$0.014$\pm$0.099 &                 &  0.021$\pm$0.043 / 0.150$\pm$0.037   &                        &  0.186$\pm$0.067                        &                                        &  0.893$\pm$0.184                   &  \\
\hline
\# 5      & 0.177$\pm$0.119   & 0.186$\pm$0.105& 0.157$\pm$0.014 / 0.103$\pm$0.012    &$-$10.0$_{-1.7}^{+1.8}$ / $-$2.1$_{-1.9}^{+2.0}$ &  $-$0.0098$\pm$0.0195                   & 14.5$_{-3.3}^{+2.9}$                   &  $-$0.190$\pm$0.037                &  22.4$_{-3.7}^{+3.0}$            \\
$+$14.5 d & 0.056$\pm$0.102   &                 & $-$0.362$\pm$0.071 / $-$0.073$\pm$0.069 &                        &  0.555$\pm$0.143                        &                                        &  0.991$\pm$0.228                   &  \\\hline
\end{tabular}\\
{$^*$Dominant axes fitted excluding the \ion{Si}{{\sc II}} $\lambda$6355 and \ion{Ca}{{\sc II}}\,NIR3 features.} \\
{$^a$Relative to the estimated peak on UT 2019-01-26.7/MJD 58509.7.} \\
\end{scriptsize}
\end{table*}

\bibliographystyle{mnras}
%\bibliography{bibtex.bib,bibtex_pah.bib} % if your bibtex file is called example.bib
\input main.bbl
%%%%%%%%%%%%%%%%% APPENDICES %%%%%%%%%%%%%%%%%%%%%

\appendix

\section{Explosion Time of SN\,2019\lowercase{np}}~\label{sec_earlylc}

\begin{figure*}
\includegraphics[width=1.0\linewidth]{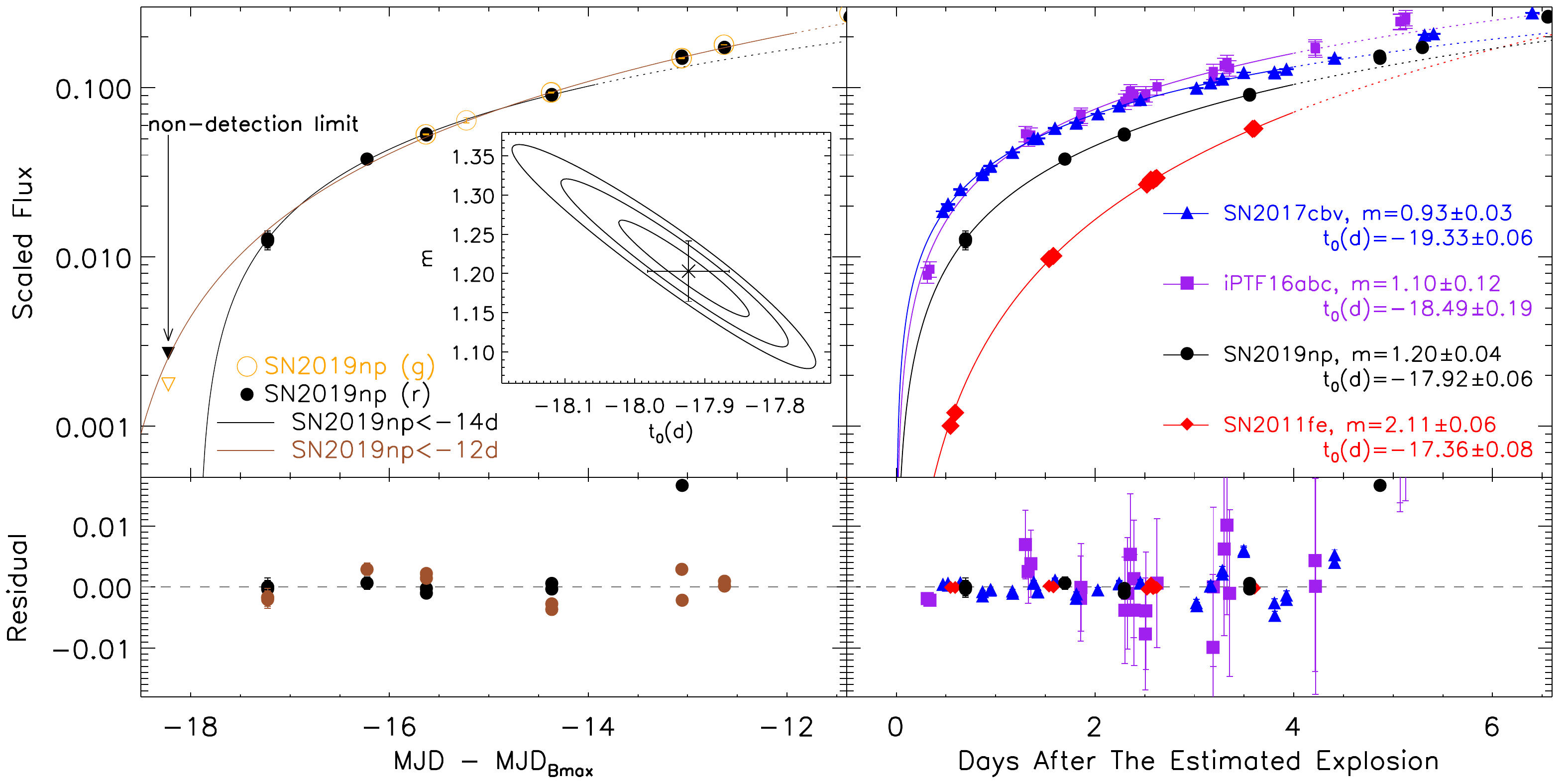}
\caption{
Best-fit $f\propto (t-t_{0})^m$ model to describe the early flux evolution of SN\,2019np compared to that of selected other SNe with well-sampled early photometry.  All flux distributions are normalised to the peak magnitude measured for each SN in the given bandpasses.  In the upper-left panel, the black and the brown solid lines fit the $r$-band flux (filled-black circles) of SN\,2019np before $-$14 and $-$12 days relative to the $B$-band maximum on MJD 58509.7, respectively.  In the inset, the inner to outer contours represent the $1\sigma$, $2\sigma$, and $3\sigma$ confidence levels of the power-law parameters.  The open orange circles mark the $g$-band photometry of SN\,2019np.  The residuals of the fits to the $r$-band light curves before $-$14 and $-$12 days are shown by the black and brown dots, respectively, in the bottom-left panel.  The upper-right panel compares the fit of SN\,2019np to the $g$ light curves of SNe\,2017cbv, 2011fe, and iPTF16abc within the first four days after explosion.  SN\,2019np exhibits a similar power-law index as SN\,2017cbv and iPTF16abc, for which a blue excess has been identified within the first $\sim 5$ days.  The residuals are shown in the bottom-right panel. 
~\label{Fig_earlylc}
}
\end{figure*}

In order to estimate the phases of the FORS2 observations relative to the explosion time of SN\,2019np, we modeled the early flux of SN\,2019np as a function of time by fitting its $r$-band flux with a power law, 
\begin{equation}~\label{Eqn_earlylc}
f(t)\propto(t-t_{0})^m\, ,
\end{equation}
where $t_{0}$ denotes the time of the first light from the explosion.  This formalism assumes that the WD exploded as an expanding fireball with constant temperature and velocity \citep{Riess_etal_1999}.  The earliest photometric dataset of SN\,2019np consists of the SDSS $g$- and $r$-band light curves generated by the \textcolor{black}{Zwicky Transient Facility} (ZTF) alert packets \citep{Patterson_etal_2019}.  \textcolor{black}{Owing} to the lack of $g$-band photometry at the earliest phases of SN\,2019np, we fitted the power law to the $r$-band flux, for which we considered two subsets, namely observations before days $\sim -14$ and $\sim -12$, respectively.  The fitting is also constrained by the last nondetection on day $-$18.2.  As shown by the filled black and open orange circles in the upper-left panel of Figure~\ref{Fig_earlylc}, after arbitrarily scaling the $g$ and $r$ light curves of SN\,2019np, the evolution of the flux in both bandpasses is consistent between days $-$15.7 and $-$12.7, when photometry is available for both filters.  The best fit to the observations before day $-$14 gives $m = 1.20 \pm 0.04$ and a rise time $t_{0} = -17.92 \pm 0.06$ days.  The statistical error of 0.06~day in $t_0$ is much smaller than the systematic uncertainty of the time of the $B$-band light-curve maximum, which amounts to $\pm$0.51 day in our analysis and has to be added.

Figure~\ref{Fig_earlylc} also compares the flux evolution of SN\,2019np to its fit with Equation~\ref{Eqn_earlylc} and includes data for selected other SNe with well-sampled photometry at similarly early phases.  The $g$-band light curve of the normal Type Ia SN\,2011fe is well approximated by an expanding fireball model (i.e., $m \approx 2$; \citealp{Nugent_etal_2011}).  Two cases with an early flux excess, namely SN\,2017cbv \citep{Hosseinzadeh_etal_2017} and iPTF16abc \citep{Miller_etal_2018}, exhibit a fast rise within the first $\sim 5$ days after the explosion and favour a power-law index around unity.  
At $m = 1.20 \pm 0.04$, SN\,2019np is similar to both of them.

\newpage
\vfill\eject
\label{lastpage}
\end{document}